\documentclass[superscriptaddress,prb,twocolumn]{revtex4-2}

\usepackage{graphicx}
\usepackage{float}
\usepackage{color}
\usepackage{amsmath}
\usepackage{amssymb}
\usepackage{natbib}
\usepackage[colorlinks = true,
            linkcolor = blue,
            urlcolor  = blue,
            citecolor = blue,
            anchorcolor = blue]{hyperref}
\usepackage[pagewise,modulo]{lineno}
\usepackage{academicons}
\usepackage{tikz,xcolor,hyperref}
\usepackage{etoolbox}

\newrobustcmd*{\mycircle}[1]{\tikz{\filldraw[draw=#1,fill=#1] (0,0) circle [radius=0.07cm];}}
\newrobustcmd*{\myholowcircle}[1]{\tikz{\filldraw[draw=#1,fill=white] (0,0) circle [radius=0.07cm];}}
\newrobustcmd*{\myholowsquare}[1]{\tikz{\filldraw[draw=#1,fill=white] (0,0) rectangle ++(4pt,4pt);}}
\newrobustcmd*{\mytriangle}[1]{\tikz{\filldraw[draw=#1,fill=#1] (0,0) --(4pt,0) -- (2pt,4pt);}}
\newrobustcmd*{\myholowtriangle}[1]{\tikz{\filldraw[draw=#1,fill=white] (0,0) --(4pt,0) -- (2pt,4pt);}}
\newrobustcmd*{\mydowntriangle}[1]{\tikz{\filldraw[draw=#1,fill=#1] (-2pt,4pt) --(0pt,0pt) -- (2pt,4pt);}}
\newrobustcmd*{\myholowdowntriangle}[1]{\tikz{\filldraw[draw=#1,fill=white] (-2pt,4pt) --(0pt,0pt) -- (2pt,4pt);}}

\begin{document}

    \title{Diffraction of fast heavy noble gas atoms, Ar, Kr and Xe on a LiF(001) surface\\Changing the tip of a 'perfect' AFM}
    
    \author{M. Debiossac}
    \affiliation{Institut des Sciences Mol\'{e}culaires d'Orsay (ISMO), CNRS-Universit\'{e} Paris-Saclay, Orsay, France}
    \affiliation{German Aerospace Center (DLR), Institute of Quantum Technologies, Wilhelm-Runge-Straße 10, Ulm, Germany}
    
    \author{P. Pan}
    \affiliation{Institut des Sciences Mol\'{e}culaires d'Orsay (ISMO), CNRS-Universit\'{e} Paris-Saclay, Orsay, France}
    
    \author{C. Kanitz}
    \affiliation{Institut des Sciences Mol\'{e}culaires d'Orsay (ISMO), CNRS-Universit\'{e} Paris-Saclay, Orsay, France}
    \affiliation{German Aerospace Center (DLR), Institute of Quantum Technologies, Wilhelm-Runge-Straße 10, Ulm, Germany}
    
    \author{P. Roncin}
    \affiliation{Institut des Sciences Mol\'{e}culaires d'Orsay (ISMO), CNRS-Universit\'{e} Paris-Saclay, Orsay, France}
    
    \date{\today}

    \begin{abstract}
    We investigate experimentally the diffraction of fast atoms of noble gas on a LiF(100) crystal oriented along the [100] and [110] directions. The wavelengths are so short that the observed quantum features are qualitatively described by semi-classical models. 
    With increasing mass and energy, the scattering profiles show an increasing number of diffraction peaks forming an increasing number of supernumerary rainbow peaks but progressively weakening in contrast. The innermost peaks corresponding to individual Bragg peaks disappear first. Along the [100] direction, only one type of atomic row contributes to the diffraction signal. After removing the contributions of the attractive forces, we present topological corrugation that should compare with those accessible with an atomic force microscope (AFM).

    \end{abstract}

    \maketitle
\section{Introduction}


The diffraction of fast atoms on crystal surfaces at grazing angle of incidence (GIFAD) has emerged in the last decades as a new tool to investigate pristine crystal surfaces and to track the growth of thin films.
It uses atoms with a kinetic energy $E_0\sim$ keV and the same grazing geometry as reflection high energy electrons diffraction, leaving the volume above the surface free for evaporation cells.
In this grazing geometry, the perfect decoupling of the fast motion parallel to the probed crystal axis and the slower one perpendicular to it has been demonstrated for elastic diffraction \cite{Zugarramurdi_2012,Pollak2024GIFAD,Roncin_PRB_2017}. 
By tuning the angle of incidence $\theta_i$ between typically 0.2$^\circ$ and 2$^\circ$, the effective interaction energy $E_\perp=E_0\sin^2\theta_i$ can be adjusted from a few meV to several eV. 
In the elastic regime, GIFAD is equivalent to thermal energy helium scattering (TEAS) but it differs in the inelastic regime because the momentum transfer to the surface atoms occurs in several gentle collisions so that elastic diffraction can be observed at larger values of $E_\perp$, higher surface temperature and with heavier projectiles:
We investigate here experimentally and via simulations the consequences of increasing the mass of the projectile atoms.
\begin{itemize}
    \item For comparable energy the wavelength reduces (see Table \ref{tab:Wavelength}) allowing potentially higher spatial resolution but the Bragg angle $\phi_B$ separating diffraction spots reduces with $\lambda$ requiring more collimated beams.
    \item The momentum transferred to the surface increases rapidly, increasing significantly the probability of phonon excitation as described by the Debye-Waller factor adapted to GIFAD~\cite{Rousseau_2008,Manson_PRB_2008,pan2021polar}.
    \item The larger number of valence electrons increases the magnitude of the Pauli repulsion, pushing the classical turning point away from the surface. This is balanced by the increased polarizability attracting the projectile towards the surface.
\end{itemize}
Within the Born-Oppenheimer approximation, the diffracted intensities correspond to the quantum scattering of the projectile atom in the potential energy landscape (PEL) describing the energy of the projectile at any location $E(x,y,z)$ above a surface lattice unit. 
A complete approach such as \textit{e.g.} wavepacket propagation \cite{Rousseau_2007,Schuller2010,Diaz_2016b} or close-coupling \cite{zugarramurdi2013surface} provides exact correspondence between the PEL and the diffraction patterns at various energies, as well as possible bound state resonances~\cite{jardine2004ultrahigh,Debiossac_PRL_2014} or quantum reflection~\cite{zhao2008quantum}. This is specific to the quantum regime where $\lambda$ is larger than typical dimensions of the surface topology. 
The first direct consequence of the reduced wavelength attached to larger masses is that the observations presented here correspond to the semi-classical regime and can, at least qualitatively, be explained by phases attached to classical trajectories and interferences between different paths ending at identical final scattering angles $\theta_f$, $\phi_f$. 
In this regime the connection to the surface topology is more direct and the information derived should compare with that from atomic force microscopy (AFM). For example, the corrugation amplitude describing the size of bumps measured by an AFM on top of the surface is also measured by GIFAD but in the reciprocal space.
Such analogy was popularized by the semi-quantitative hard corrugated wall model (HCW). It is an optical model considering straight line reflection from the iso-energy surface $z_E(x,y)$ assumed to represent the surface where the classical trajectories are reflected. The diffracted intensities are therefore obtained by a Fourier transform of the topology.
However, as a reciprocal space technique, atomic diffraction measures scattering angles acquired essentially while bouncing off these bumps. However, trajectories are also affected by angular changes due to the attractive forces modifying outgoing trajectories.

The vast majority of publications on atomic diffraction on surfaces consider He as a projectile (see \textit{e.g.} Ref.~\cite{holst2021material} for a recent review at thermal energies) but numerous theoretical studies~\cite{legrand1986nearly,moix2011communication,gravielle2013interaction,miraglia2017reexamination} as well as a few experimental works considering Ne~\cite{mattera1983quasi,rieder1984observation,semerad1987resonant,minniti2012helium,debiossac2020refraction} have been published.

Detailed data for helium~\cite{Winter_PSS_2011,debiossac2021grazing} and neon~\cite{Gravielle_2011,debiossac2020refraction,debiossac2023elastic} have already been published over a wide range of energy and angle and only the new results with Ar, Kr and Xe will be presented. 
Rather than three independent section with each projectile, the paper focuses on specific topics and their evolution along the Ar, Kr and Xe sequence.
The outline is as follows: 
After a brief recall of the GIFAD technique in Sec. \ref{ch:setup}, Sec. \ref{ch:elastic} to  Sec. \ref{ch:classical} are a purely experimental description of the evolution from quantum to classical features. 
The elastic and inelastic diffraction peaks are presented in Sec. \ref{ch:elastic} and Sec. \ref{ch:inelastic} respectively. 
Section \ref{ch:supernum} presents the supernumerary rainbows while Sec. \ref{ch:rainbow} focuses on the classical rainbow and Sec. \ref{ch:classical} focuses on comparatively larger collision energy where no trace of quantum behavior can be seen but where specific features can be efficiently described by semi-classical models.

Section \ref{ch:analysis} presents all the data in a more comprehensive way highlighting the role of the attractive forces extending up to eV effective energies and that of the topology, dominant at larger energies.  
A model potential energy landscape fitted to the diffraction data is presented for the [100] direction and the resulting surface topology is derived for all noble gases along this direction before tracing perspectives and conclusion in Sec.\ref{ch:conclusion}.

\begin{figure}[h]
    \centering
    \includegraphics[width=0.75\linewidth,trim={1.5cm 4cm 9cm 2.8cm},clip,angle =0,draft = false]{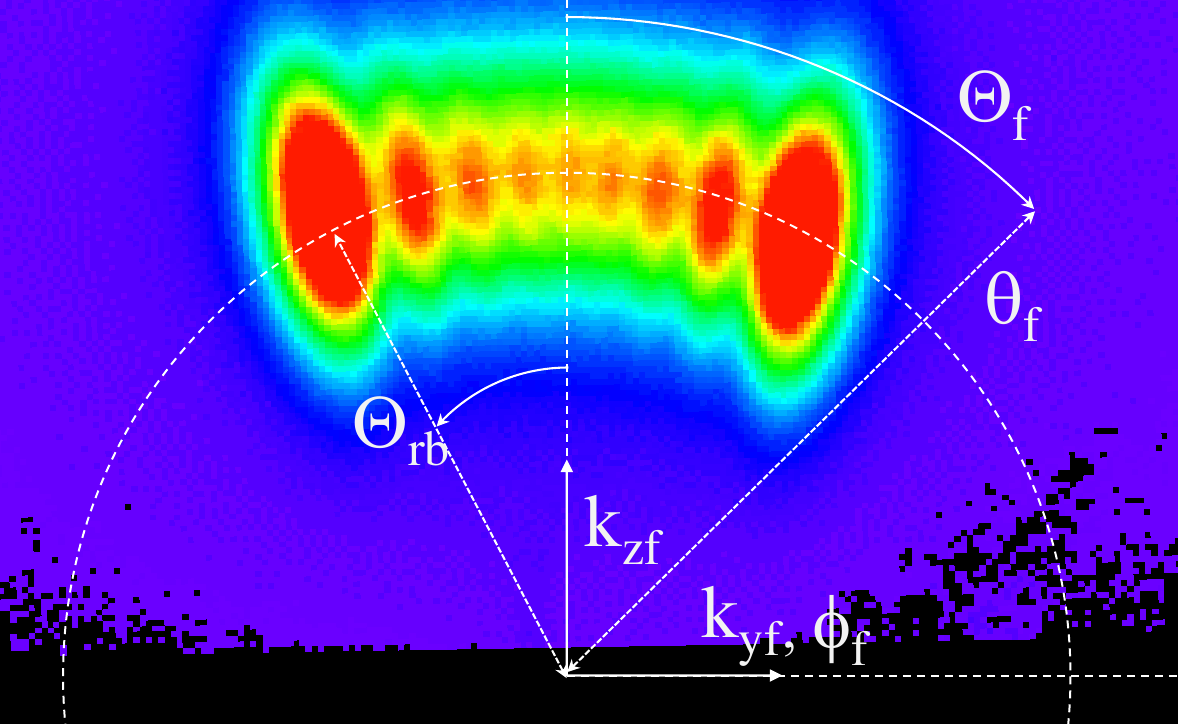}
    \caption{Raw camera capture of the scattered atoms on the detector for 500 eV Ar impinging LiF(001) along [110] direction at $\theta_i=1^{\circ}$, $\phi_i=0^{\circ}$. The dashed circle $\theta_f=\theta_i$ corresponds to energy conservation; $|k_f|=|k_i|$, the vertical line is the specular plane $k_{y_f}=k_{y_i}$. In the lab frame, the azimuthal (lateral) deflection angle is $\phi_f=\arctan(k_{y_f}/k_{x_f})\simeq k_{y_f}/|k_{i}|$, less than a degree while the angle in the $(y,z)$ detector plane is $\Theta_f=\arctan({k_{y_f}/k_{z_f}})\simeq \arctan(\phi_f/\theta_i)$ is equivalent to the angles measured in TEAS and can reach $\pm 90^\circ$.}
    \label{fgr:Fig1}
\end{figure}

\section{Experimental setup}\label{ch:setup}
A detailed description of the GIFAD setup and of the associated coordinate system can be found in Refs.~\cite{pan2022setup} and ~\cite{Debiossac_Nim_2016} respectively and only a simple sketch is presented below.
An ion beam of kinetic energy $E_0$ is neutralized in a charge exchange cell and severely collimated by two diaphragms half a meter apart to reduce the angular divergence to the 0.1 mrad range.
It impinges the crystal surface at a grazing angle of incidence $\theta_i \sim\;1^\circ$, and all the diffracted beams are recorded simultaneously on a microchannel plate imaging detector~\cite{lupone2018large}.

Figure~\ref{fgr:Fig1} displays a scattering pattern typical of short projectile wavelength where mainly supernumerary rainbow structures are visible as discussed in Sec.\ref{ch:supernum}. 
It also illustrates the angular coordinates used here, $\theta$ and $\phi$ are referred to as the polar and azimuthal angles. 
The azimuthal (lateral) deflection will be reported in degree or in multiple values of the Bragg angle. When comparing different energies or different incidence angles we use the relative angle $\Theta_f=\arctan{k_{y_f}/k_{z_f}} \equiv \arctan{\phi_{f}/\theta_{i}}$ to suppress the trivial dependence of the  azimuthal scattering $\phi_f$ with the angle of incidence $\theta_i$. This then compares directly with the coordinates of a TEAS setup where the incidence angle is referred to the surface normal. 
The coordinate $\Theta$ is also more appropriate to describe the refraction effect relating the deflection angle $\Theta_{\text{obs}}$ observed on the detector  to the one $\Theta_{\text{surf}}$ acquired when bouncing on the bumps, using the simple Snell-Descartes formulation: 
\begin{equation}
\sin\Theta_{\text{obs}} \sqrt{E_\perp}=\sin\Theta_{\text{surf}} \sqrt{E_\perp+D}
\label{eq:Snell}        
\end{equation}
where $D$ is the well-depth resulting from attractive forces so that $E_\perp+D$ represents the effective energy when hitting the surface.

\section{Tracking the elastic diffraction}\label{ch:elastic}
Elastic diffraction is identified as a narrow peak at the specular angle in the polar scattering profile (see e.g. Fig. 3 in Ref.~\cite{pan2021polar}). 
Each elastic diffraction spot has a profile very close to that of the primary beam, the latter being limited by the size of diaphragms and detector resolution. 
For He and Ne projectiles on the LiF(001) surface, the ratio of elastic to inelastic diffraction was found independent on the crystal direction and could be modeled by the Debye-Waller factor adapted to GIFAD~\cite{Rousseau_2008,Manson_PRB_2008}. 
At room temperature the following scaling was observed~\cite{pan2022temperature} DWF=$ A(M_p) e^{-\frac{1}{8}M_p E_0 \theta_i^3}$ with $M_p$ the projectile mass in Dalton, $E_0$ its total kinetic energy in meV and $\theta_i$ the incidence angle. For $E_0=1$ keV and $\theta=0.5^\circ$, the exponential term amounts to 70, 19, 3.6, 0.08 and 0.002 \% for He, Ne, Ar, Kr and Xe respectively. 
On top of this formula, an empirical pre-factor $A(M_p)$ of 0.55 for He and 0.16 for Ne was observed \cite{pan2021polar} probably due to our limited surface coherence and indicating that the sensitivity to these defects increases with the projectile mass.
When analyzing the polar profile associated with the file in Fig.~\ref{fgr:Ar_[100]_500eV}, we measured a very weak elastic contribution representing only 0.3$\pm0.3$ \% of the scattered intensity, at the limit of the present statistics but indicating a pre-factor below 0.1 for Ar on our LiF crystal, consistent with the lighter projectiles \cite{pan2022temperature}.
This suggests that unambiguous elastic diffraction could probably be observed at lower surface temperature or with better prepared samples.

\begin{table}  
    \centering
    \begin{tabular}{c|c|c|c|c|c|}
        Energy & He & Ne & Ar & Kr & Xe\\
        \hline
        1 keV & 0.45 & 0.20 & 0.14 & 0.098 & 0.079\\
        \hline
        100 meV & 45 & 20 & 14 & 9.8 & 7.9\\
        \hline
        10 meV & 144 & 64 & 45 & 31 & 25\\
        \hline

    \end{tabular}
    \caption{De Broglie Wavelength $\lambda$ in pm for different energies and atoms. On LiF(001), the distance $a_\perp$ between identical atomic rows along [100] and [110] are 202 and 286 pm respectively corresponding to reciprocal lattice vectors $G_\perp$=3.11 and 2.20 \AA$^{-1}$. The associated Bragg angle is $\phi_B\simeq \lambda/a_\perp$. }
    \label{tab:Wavelength}
\end{table}

\section{Tracking inelastic diffraction peaks} \label{ch:inelastic}
Along the azimuthal direction, inelastic peaks are broader than the elastic one but can be well identified as long as their line-width $\sigma_{lw}$ is less than the peak separation \textit{i.e}. the Bragg angle $\phi_B$.
 
\subsection{The inelastic azimuthal line profile}
As proposed in Refs.~\cite{Seifert_2015,pan2023lateral}, the inelastic azimuthal line profile does not depend on the crystal orientation and can easily be measured by orienting the surface far away from a low index direction also called random direction [Rnd] where only one diffraction peak is present. 
This was investigated in detail in the quasi elastic regime with He and Ne projectiles where a quasi-Lorentzian line shape with reduced wings was proposed~\cite{pan2023lateral} in the form of the product of a Lorentzian by a Gaussian 
\begin{equation} \label{eq:LG} 
LG(\phi)=A \cdot e^{-\phi^2/2w^2}/(4w^2+\phi^2),
\end{equation}
where the standard deviation $\sigma_{lw}$ is related to the width parameter $w$ by $\sigma_{lw}\simeq 0.7 w$ and A is a normalization factor. All the line shapes used here and plotted in green in Figs.~\ref{fgr:Ar_[100]_500eV}-\ref{fgr:Ar_1784_s=4}, use this form  with a width parameter interpolated from measurement along the [Rnd] direction~\cite{sigma_phi} and referred to by its standard deviation $\sigma_{lw}$.

\subsection{The [100] direction}

\begin{figure}[h]
\includegraphics[width=0.75\linewidth,trim={0 0 0 0},clip,angle =0,draft = false]{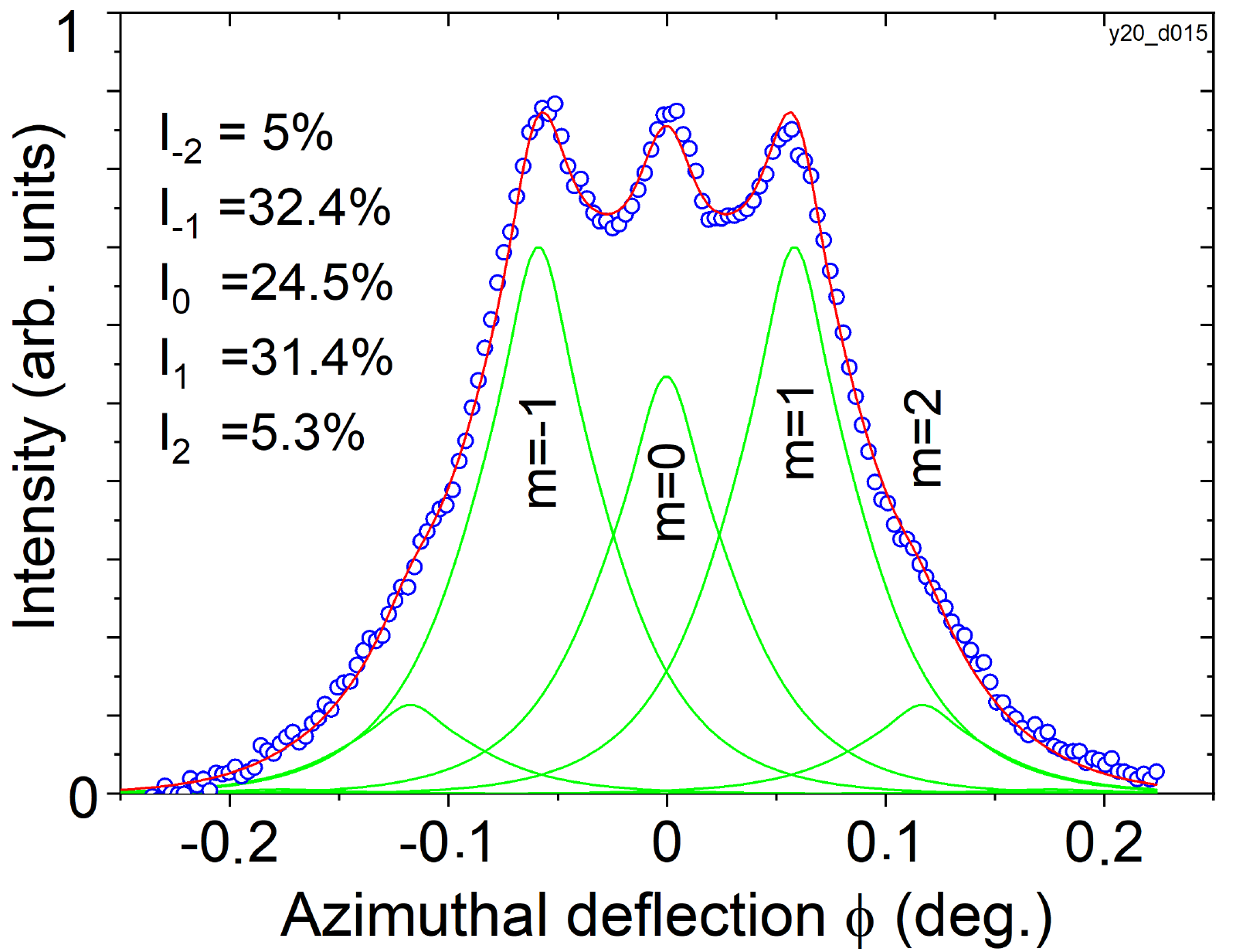}
\caption{\label{fgr:Ar_[100]_500eV} For 500 eV Ar impinging LiF along [100] at $\theta_i$=0.29$^\circ$, the diffracted intensity measured on the specular circle is fitted (red line) by quasi Lorentz line-shape (eq.\ref{eq:LG}) measured along the [Rnd] direction. The measured intensities are reported on the inset. 
}\end{figure}

Figure~\ref{fgr:Ar_[100]_500eV} shows an azimuthal scattering pattern corresponding to a narrow band around the specular circle~\cite{Specular} for 500 eV Ar atoms. 
The inelastic peaks are resolved but overlap significantly. 
Using the line profile recorded along the [Rnd] direction, the only free parameters of the fit are the diffracted intensities $I_m$ reported in the left-hand side. 
These intensities are very close to the values $I_m=J_m^2(\zeta=1.52)$ where $J_m$ are the Bessel functions of order $m$ so that the full diffraction pattern can be fitted by a single number $\zeta$, interpreted in the following. 
This corresponds to the predictions of semi-classical perturbation theory \cite{Pollak2024GIFAD} as well as that of the hard corrugated wall model (HCW). In the latter, the PEL above the surface is represented by equipotential lines with a pure sinusoidal shape $z_E(y)=\frac{z_c}{2} \cos (G_\perp y)$ providing a simple interpretation for $\zeta=k_\perp z_c$ and $z_c$ is called the corrugation amplitude. 
For instance the value of $\zeta=1.52$ derived above translates into $z_c\sim$ 10 pm for the associated wave vector $k_\perp$=15.6 \AA$^{-1}$. We will conclude in Sec.\ref{ch:analysis} that this value is almost twice too large. However, interpreting $\zeta$ in terms of classical rainbow angle gives a correct value even if this latter is somewhat hidden in the data of Fig.~\ref{fgr:Ar_[100]_500eV}. 

\begin{equation} \label{eq:rb} 
\Theta_{rb}=-2\arctan\left(\zeta\frac{G_\perp}{2 k_\perp}\right)\simeq \zeta\frac{G_\perp}{k_\perp} =\zeta\frac{\lambda_\perp}{a_\perp} 
\end{equation}

Along the [100] direction, the same agreement with Bessel functions was observed for He and Ne projectiles (see Fig. 3 of Ref.\cite{Rousseau_2007}, Fig. 3.12 of Ref.~\cite{Winter_PSS_2011} and Fig.9 of Ref.~\cite{debiossac2023elastic}). 

\begin{figure}[h]
    \centering
    \includegraphics[width=0.9\linewidth,trim={0cm 0cm 0cm 2cm},clip]{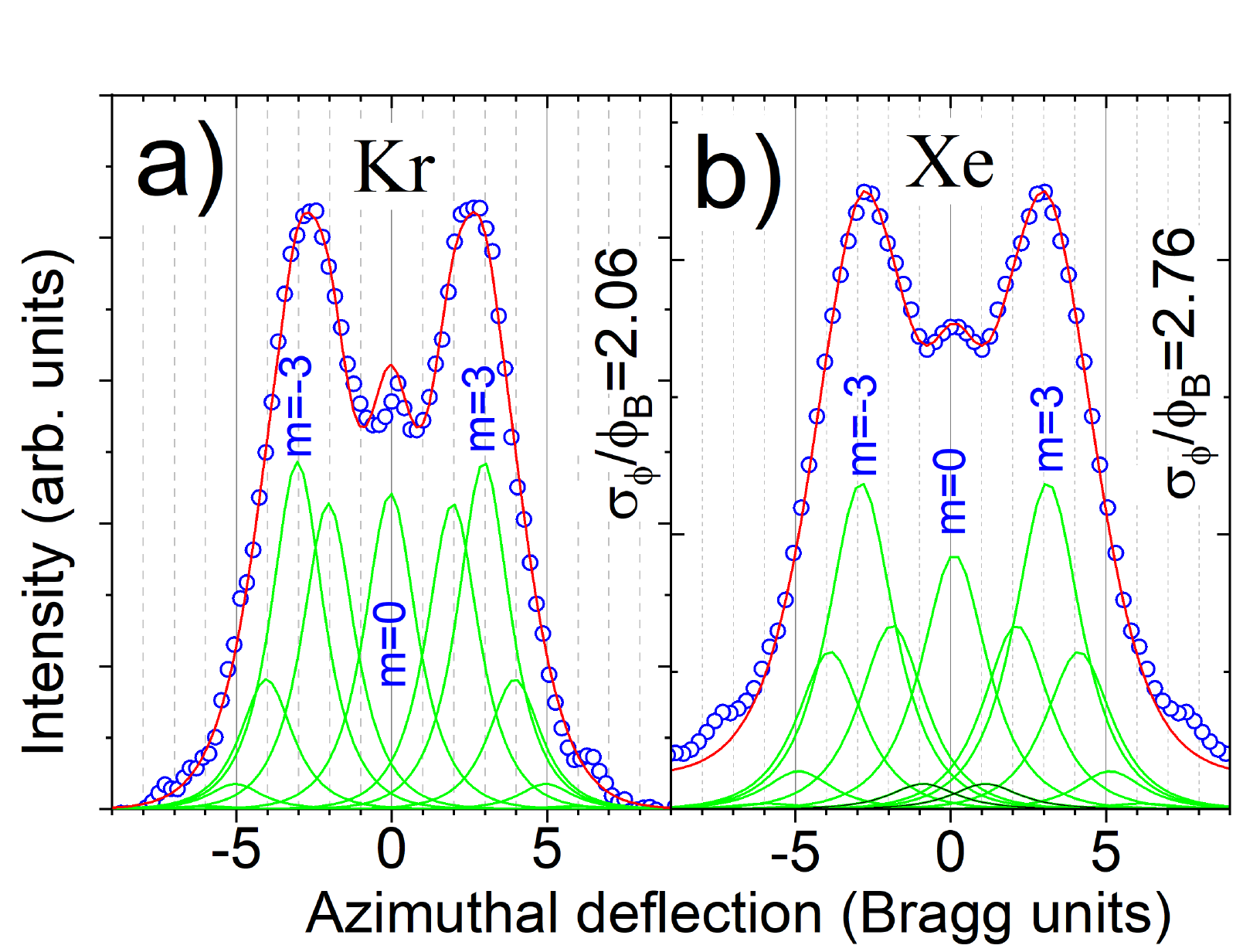}
    \caption{Along the [100] direction, scattering profiles of 4 keV Kr and Xe atom at $\theta_i=0.45^\circ$ and $0.48^\circ$ respectively, have been fitted by a HCW formula giving comparable values of $\zeta$ close to 3.9 for which the intensity $I_{\pm 1}=J_1^2(\zeta)$ is close to zero (see text). Data for Xe have been symmetrized.}
    \label{fgr:Kr_Xe_Fit_HW}
\end{figure}
For Kr and Xe, we measured, along the [Rnd] direction, an inelastic width $\sigma_{lw}$ larger than the Bragg angle and, consistently, we could not clearly resolve adjacent diffraction peaks.
However, one can easily identify situations where the $m=\pm1$ diffraction peaks are weak because their intensity oscillate in quadrature with $m=0$ and $m=\pm2$, offering the opportunity to identify diffraction peaks in spite of $\sigma_{lw}>\phi_B$.
The first zero in the Bessel function $J_1(\zeta)$ occurs for $\zeta$ = $\pi+\pi/4\sim$ 3.9 where the $\pi/4$ is interpreted as the Gouy phase or Masslov index 'naturally' present in the Bessel function. 
The scattering profiles displayed in Fig.~\ref{fgr:Kr_Xe_Fit_HW} correspond to this criterion. 

\subsection{The [110] direction}
\begin{figure}[h]
\includegraphics[width=0.85\linewidth,trim={0 0 0 0cm},clip,angle =0,draft = false]{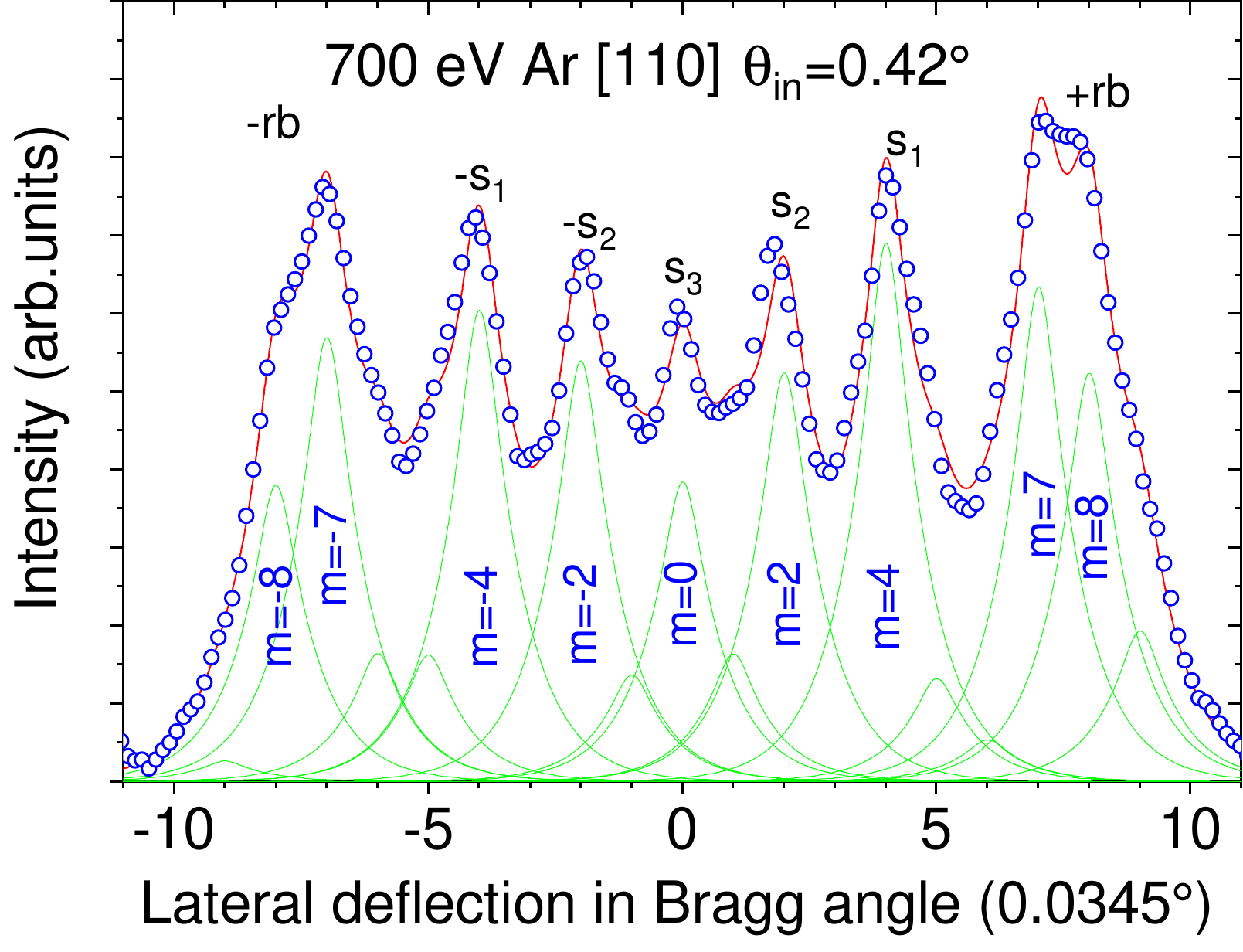}
\caption{\label{fgr:700eV_Ar_[110]} Scattering profile of 700 eV Ar impinging LiF along [110] at $\theta_i=0.42^\circ$ ($E_\perp$=38 meV). The inelastic diffracted intensity is fitted (red line) by line-shape measured along a [Rnd] direction. The peaks labeled s$_i$ correspond essentially to even diffraction orders $|m|$=0,2,4 while $|m|$=7, 8, 9 contribute to the outer classical rainbow.}
\end{figure}

Figure~\ref{fgr:700eV_Ar_[110]} shows a scattering profile recorded of Ar atoms along the [110] direction, together with a free fit using a profile fitted along the [Rnd] direction. 
Peaks are identified mainly because the intensity of the even and odd diffraction orders oscillate in quadrature in the paraxial region of small deflection \cite{Roncin_PRB_2017} (see \textit{e.g.} Figs 2c and 3c of Ref.~\cite{pan2021polar} for illustration).

Along this direction, as already noticed for He~\cite{momeni2010grazing} and Ne~\cite{debiossac2023elastic} projectiles, the peak intensities $I_m$ are poorly described by the HCW model using a pure cosine making the HCW predictions less analytic.

\section{Tracking the supernumerary rainbows}\label{ch:supernum}

Another direct evidence of quantum behavior is the presence of supernumerary rainbows. 
Their observation in natural conditions was essential in the foundation of the wave nature of light~\cite{young2016bakerian} and are quantitatively described by the Mie theory of light scattering\cite{laven2004simulation}.
More generally, supernumerary rainbows are ubiquitous in quantum scattering theory. 
A classical rainbow is associated with an extrema in the deflection function and, most often, on each side of this extrema, two different semi-classical trajectories are scattered at the same final angle and may interfere. 
The associated phase shift $\zeta$ between these trajectories increases as these trajectories separate and supernumerary rainbows are observed when the phase shift reaches multiple of $2\pi$. 
Therefore, supernumerary rainbows indicate a situation with $\lambda\gg z_c$  where pure quantum behavior, such as bound state resonance~\cite{chow1976bound,Debiossac_PRL_2014} or quantum reflection~\cite{zhao2008quantum,miret2017scattering} can probably be neglected. 
In GIFAD, increasing the number of diffraction orders rapidly leads to a quasi-continuous profile \cite{garibaldi1975quantum,Schueller2008Supernumerary}, 
as if the discrete values of $J_m^2(\zeta)$ would converge to the envelop $J_\nu^2(\zeta)$ where $J_\nu$ is the uniform Bessel-Clifford function such that $J_\nu(\zeta)=J_m(\zeta)$ where $\nu=\phi_f/\phi_B$ (see \textit{e.g.} Fig. 2.5 of ref.\cite{Winter_PSS_2011}). This would be exact if the line profile were negligible, which partly contradicts the fact that individual peaks cannot be resolved. 

\subsection{Along the [110] direction}
Along the [110] direction, such scattering profiles where individual diffraction peaks seem absent have been observed with helium~\cite{Schueller2008Supernumerary,debiossac2021grazing}, neon~\cite{Gravielle_2011,debiossac2020refraction} and argon (Fig.~\ref{fgr:Fig1}). 
Two intense classical rainbows are present at the maximum lateral deflection striding over seven weaker maxima associated with four supernumerary rainbows for positive or negative deflection.

\begin{figure}
\includegraphics[width=0.8\linewidth,trim={0 0 0 0cm},clip,angle =0,draft = false]{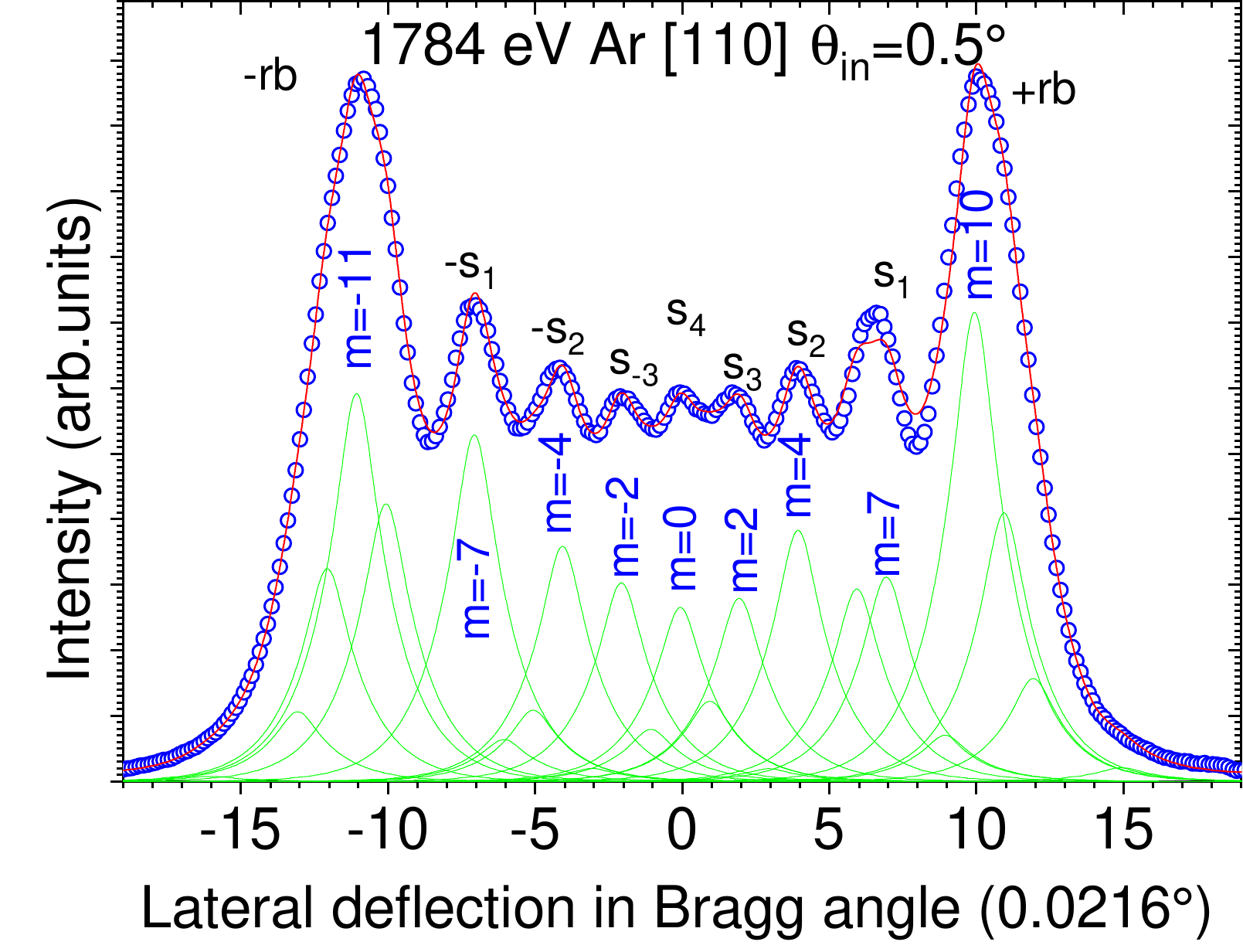}
\caption{\label{fgr:Ar_1784_s=4} Scattering profile of 1784 eV Ar impinging LiF along [110] at $\theta_i=0.5^\circ$. The red line is a fit with a line-shape width $\sigma_{lw}\approx$ 30\% larger than the Bragg angle. 
}
\end{figure}

Figure~\ref{fgr:Ar_1784_s=4} shows a similar profile in a more quantitative plot where the supernumerary rainbows are numbered $s_1$ to $s_4$ starting from the outer classical rainbow position. 
Figure~\ref{fgr:Ar_1784_s=4} also displays a free fit by individual lines showing that the inner supernumerary rainbow peaks $s_2$ to $s_4$ still correspond to individual diffraction peaks $m=0,\pm2$ and $\pm4$ separated by 2$\phi_B$ as in Fig.~\ref{fgr:700eV_Ar_[110]}.

This outlines the smooth continuity between scattering patterns with resolved inelastic peaks and supernumerary rainbows but also the associated difficulty because the position of the inner peaks is first sticking to the location of actual diffraction peaks while it evolves continuously in the Bessel-Clifford function.

\begin{figure}
\includegraphics[width=0.9\linewidth,trim={1cm 0 5.5cm 0},clip,angle =0,draft = false]{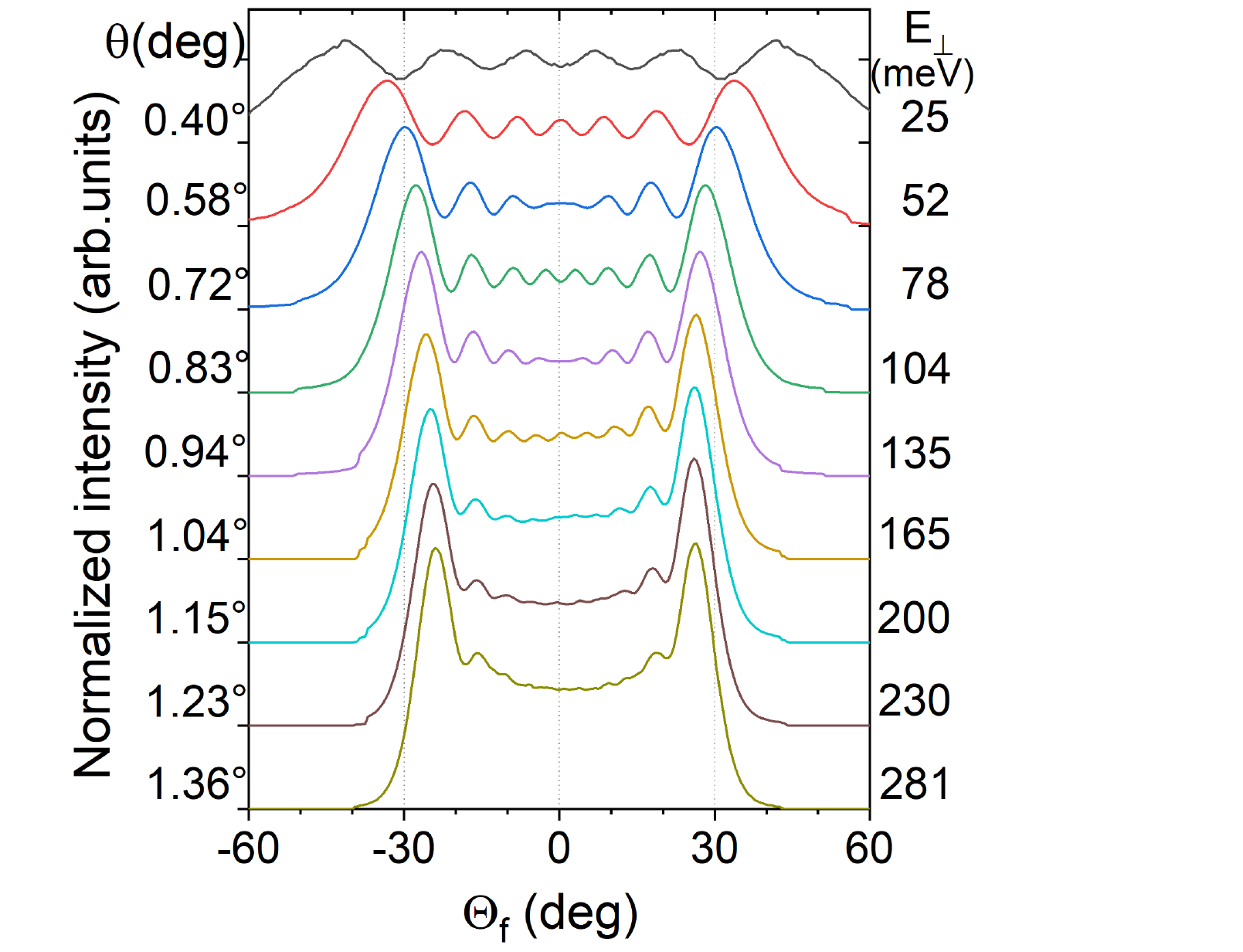}
\caption{\label{fgr:Ar_500_110_profiles} $\Theta$ scattering profiles recorded on the specular circle for 500 eV Ar along LiF[110]. The angle of incidence $\theta_i$ is indicated on the left and the associated value of $E_\perp$ is on the rhs.}
\end{figure}

Figure~\ref{fgr:Ar_500_110_profiles} reports the evolution of the scattering pattern during a $\theta$-scan where low angles and low energies are plotted on top. 
The number of supernumerary rainbows increases from two to four between $\theta_i=0.4^\circ$ and $1.04^\circ$. However, the attenuation of the contrast makes their identification more and more hazardous while the classical rainbow peak becomes rapidly dominant. 

We have also checked that the scattering pattern only depends on the energy $E_\perp$. 
The scattering patterns of 500 eV and 1789 eV recorded with $E_\perp$= 100 meV were found almost undistinguishable when plotted on a common $\Theta_f$ scale. Both display peaks associated with three super-numerary rainbows at a very comparable level of contrast. 
The strict equivalence was established for the elastic intensity but seems to apply for inelastic intensities on the specular circle $\theta_f=\theta_i$ suggesting that the inelastic intensities are probably also close to the elastic one.

\begin{figure}
\includegraphics[width=0.85\linewidth,trim={0 0 0 0},clip,angle =0,draft = false]{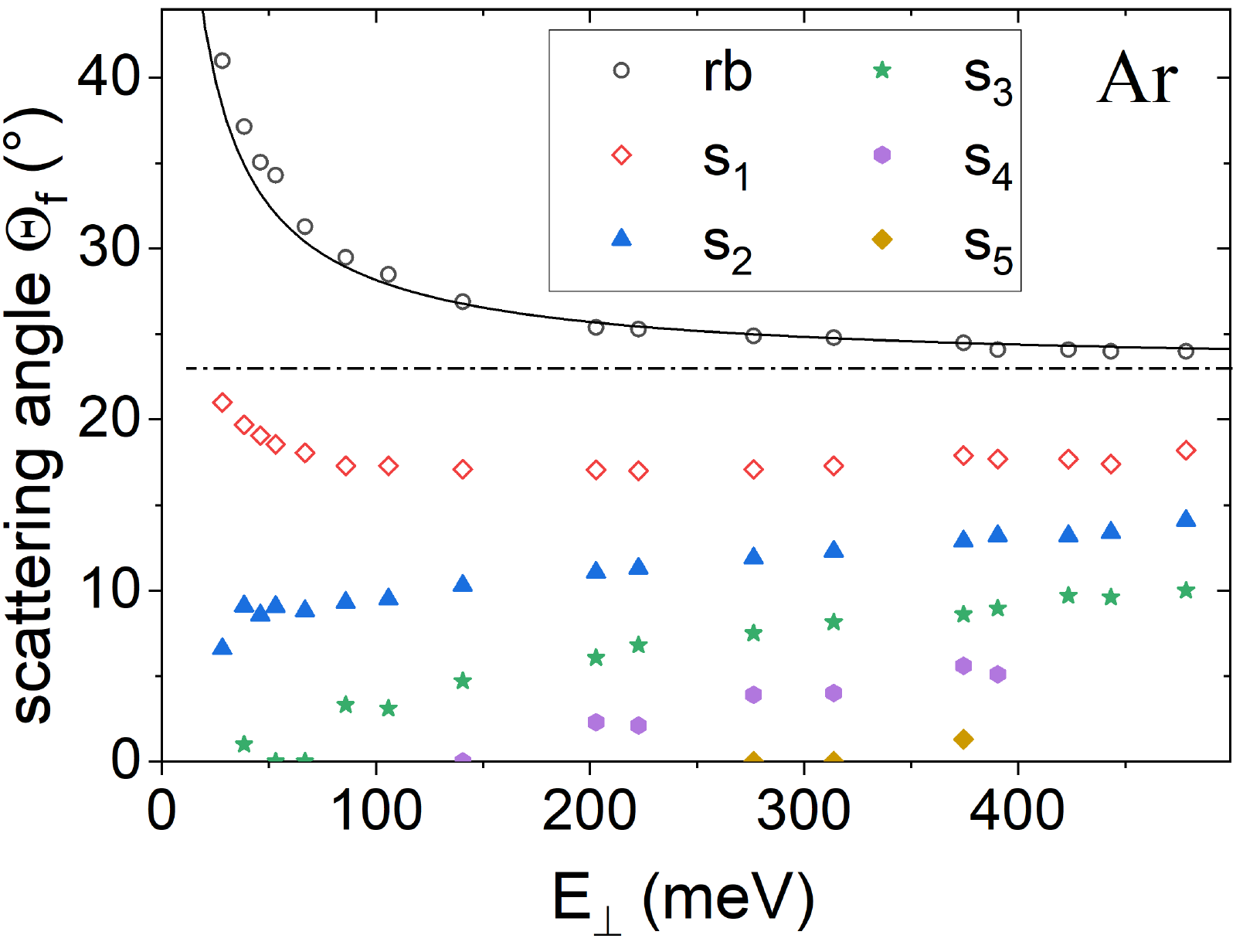}
\caption{Evolution of the rainbows peaks with $E_\perp$ for 1784 eV Ar along the [110] direction. The positions are measured by hand and the continuous black line is a Snell transform of the horizontal dashed line with $D=50$ meV.}\label{fgr:Ar_1784_110_Rb_peaks} 
\end{figure}

\begin{figure}[h]
\includegraphics[width=0.85\linewidth,trim={0 0 0 0},clip,angle =0,draft = false]{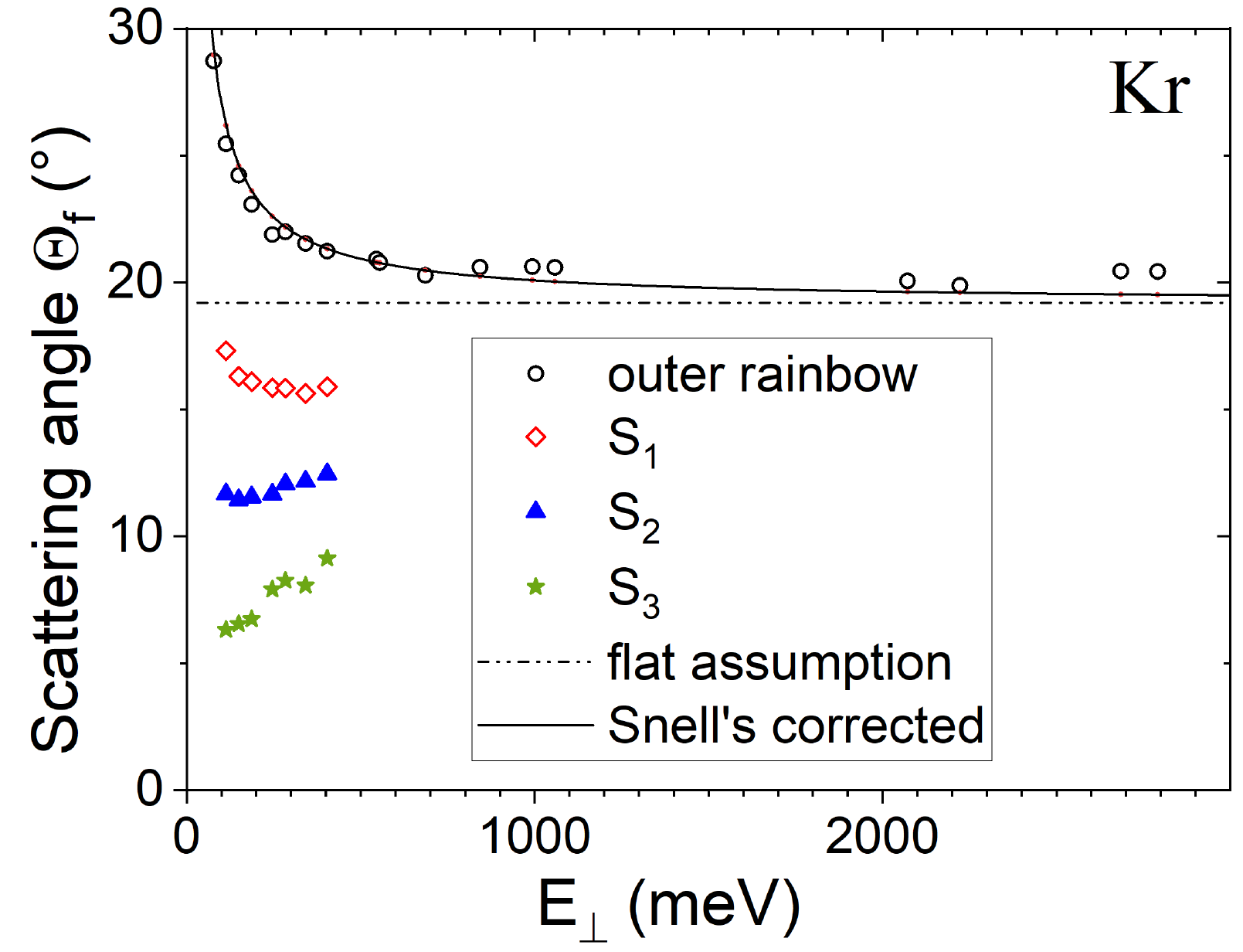}
\caption{ For 4 keV Kr atoms along the [110] direction, the peak position of the classical outer rainbow and inner supernumerary rainbows is plotted as a function of $E_\perp$. The full black line corresponds to an optical model using Snell's law with $\Theta_0$=19.2$^\circ$ and a well-depth $D$=92 meV.}\label{Kr_110_4k_rainbow}
\end{figure}

\begin{figure}
    \centering
    \includegraphics[width=0.80\linewidth]{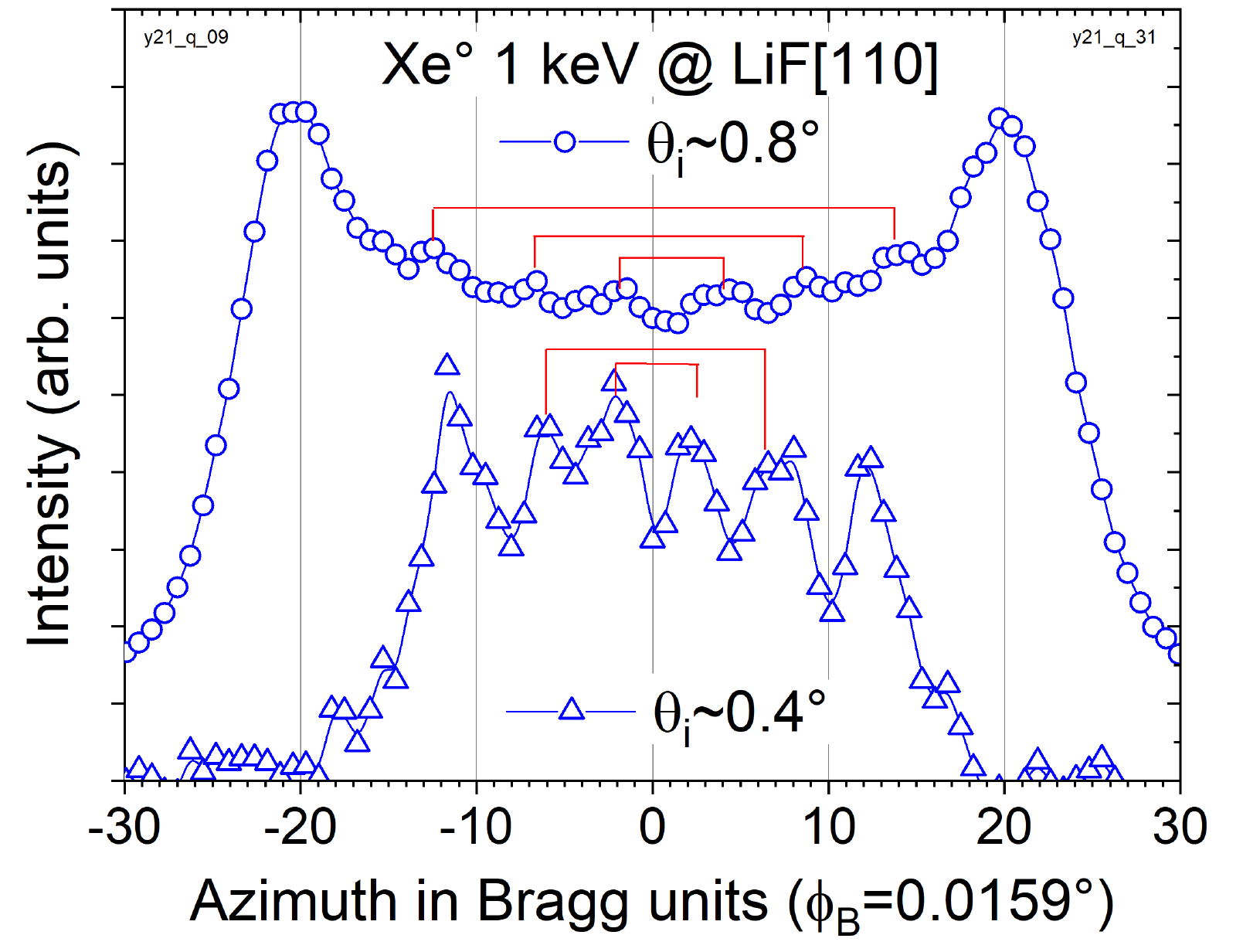}
    \caption{Azimuthal scattering profile of 1 keV Xe impinging LiF[110] at $\theta_i$=0.4$^\circ$ and 0.8$^\circ$. The statistics is poor but indications of two supernumerary rainbows are present at $\theta_i$=0.4$^\circ$ and possibly three or four at $\theta_i$=0.8$^\circ$. The full line is a B-spline through the data points.
    }
    \label{fig:Xe_[110]}
\end{figure}

Similarly to Ref.~\cite{Gravielle_2011} for Ne, the angular location of the rainbow and supernumerary rainbow peaks is reported using the angle $\Theta=\arctan{(\phi_f/\theta_i)}$, in Fig.~\ref{fgr:Ar_1784_110_Rb_peaks} for Ar and in Fig.~\ref{Kr_110_4k_rainbow} for Kr. 
For Xe, Fig.~\ref{fig:Xe_[110]} shows that supernumerary rainbows could be present but their identification is limited both by resolution and statistics.


\subsection{Along the [100] direction}
Along the [100] direction and at low value of $E_\perp$, the corrugation amplitude is typically 3-4 time smaller than along the [110] and the number of supernumerary rainbows is reduced by the same amount. 
In Fig.~\ref{fgr:Kr_Xe_Fit_HW}, the central $m=0$ peak surrounded by very weak $m=\pm1$ peaks and then dominant $m=\pm2,\pm3$ can be seen as the evidence of a single supernumerary rainbows peak $s_1$ for Kr and Xe. 
For Ne, Fig.8b) of Ref.~\cite{debiossac2023elastic} shows a better resolved but comparable profile while Fig.~\ref{fgr:Ar[100]_Im&zc}b) indicates that, for Ar, a similar situation with $I_1\simeq$ 0 is reached with $E_\perp\sim$ 250 meV. 
More supernumerary rainbows probably develop at larger energy $E_\perp$ but, at room temperature $\sigma_{lw}$ washes out the central structure where peaks are separated by $2\phi_B$.

\section{Tracking the Rainbow angle}\label{ch:rainbow}

Before the discovery of GIFAD, the rainbow angle was a reference measurement to compare theoretical PEL with scattering experiments under axial channeling conditions~\cite{schuller2005interatomic,specht2011rainbow,gravielle2013interaction} with interaction energies $E_\perp$ above a few eV. 
At these energies, the rainbow structure is the most salient feature of the scattering profile but tracking its position accurately is not straightforward. 
The rainbow has a classical origin but the exact shape is deeply affected by quantum features. 
In the quantum regime where $\lambda \eqslantgtr z_c$ as in Fig.~\ref{fgr:Ar_[100]_500eV} it does not correspond to any visible feature. 
Then, when supernumerary rainbows start to emerge, the shape of the rainbow peak becomes an Airy-like profile (see e.g. eq. 13 of Ref.~\cite{Winter_PSS_2011}) affected by the inelastic line-width $\sigma_{lw}$ associated to the thermal agitation of the surface atoms. 
At larger energy $E_\perp$ and larger values of $\sigma_{lw}$ the rainbow profile is dominated by $\sigma_{lw}$ and the point of maximum intensity is considered a fair estimate of the rainbow position. 

A simple alternative is to use the statistical width of the entire scattering profile, either from quantum point of view : $\sigma_{m}^2=\Sigma_m m^2 I_m$ \cite{corrug_100} or from the classical one $\sigma_{\phi_f}^2=\Sigma_i (\alpha\,i)^2 I_i$ where $I_i$ is the intensity measured on the channel $i$ of the histogram of scattering angles normalized to unity and $\alpha$ the angular calibration. 
Both values should be close as they are connected : $\sigma_{\phi_f}^2 \sim \phi_B^2 \sigma_{m}^2 + \sigma_{lw}^2$.
This robust and parameter-free definition allows a smooth continuity through the different regimes as shown in Fig. 13 and Fig. 21 of Ref.~\cite{debiossac2021grazing}. 

We use here the coordinate $\Theta$ and the angle $\Theta_{rb}$ which offers the double advantage that it corresponds to the deflection angle measured in TEAS and that is suppresses the linear dependence with $\theta_i$ or $k_{iz}$ (compare for instance Fig.~\ref{fgr:Kr_zc_de_Ep}b)  and Fig.~\ref{fig:Xe_4k_2D} using $\Theta$ and $\phi$ scales respectively).


\subsection{Along the [110] direction}

For all energies investigated here, the supernumerary rainbow is well developed along the [110] direction indicating $\lambda\gg z_c$ (or $\zeta\gg$1 ) and the rainbow position was estimated by hand by trying to pinpoint the maximum of the outer peaks. 
The corresponding values are plotted in Fig.~\ref{fgr:Ar_1784_110_Rb_peaks} and~ \ref{Kr_110_4k_rainbow} for Ar and Kr, respectively. They show a stable value at large value of $E_\perp$ and a rapid increase at low energy. 
In these figures, the solid black line going through the data corresponds to a refraction model (Eq.\ref{eq:Snell}) with a fixed emission angle $\Theta_{\text{surf}}$ of 19$^\circ$ and 25$^\circ$ and a well-depth D of 50 meV and 92 meV for Ar and Kr respectively.
For 1 keV Xe with $E_\perp$ between 100 and 800 meV, the rainbow peak position was found to be compatible with a form : $\Theta_{rb} \simeq 13^\circ \sqrt{\frac{E+D}{E}}$ with $D=135$ meV.

\subsection{Along the [100] direction}
The rainbow angle is more difficult to pinpoint directly along the [100] direction as it is typically $\Theta_{rb}\sim\,$6$^\circ$, three to four times smaller than along [110].
Taking advantage of the good fit with Bessel functions $J_m^2(\zeta)$ in Fig.~\ref{fgr:Ar_[100]_500eV} and Fig.~\ref{fgr:Kr_Xe_Fit_HW}, 
we use this very stable data reduction procedure to extract the number $\Theta_{rb}=\zeta \,\frac{G_\perp}{k_\perp}$ which, within the HCW is the rainbow angle as long as $\Theta\sim\tan\Theta$. 
Note that with the same cosine shape and small-angle approximations $\Theta_{rb}$ is also directly connected to the standard deviation discussed at the beginning of the section~\cite{Pollak2024GIFAD}, $\Theta_{rb}\simeq \sigma_\Theta \sqrt{2}$.

\begin{figure}
    \centering
    \includegraphics[width=0.8\linewidth,trim={0 2.5cm 0 0},clip]{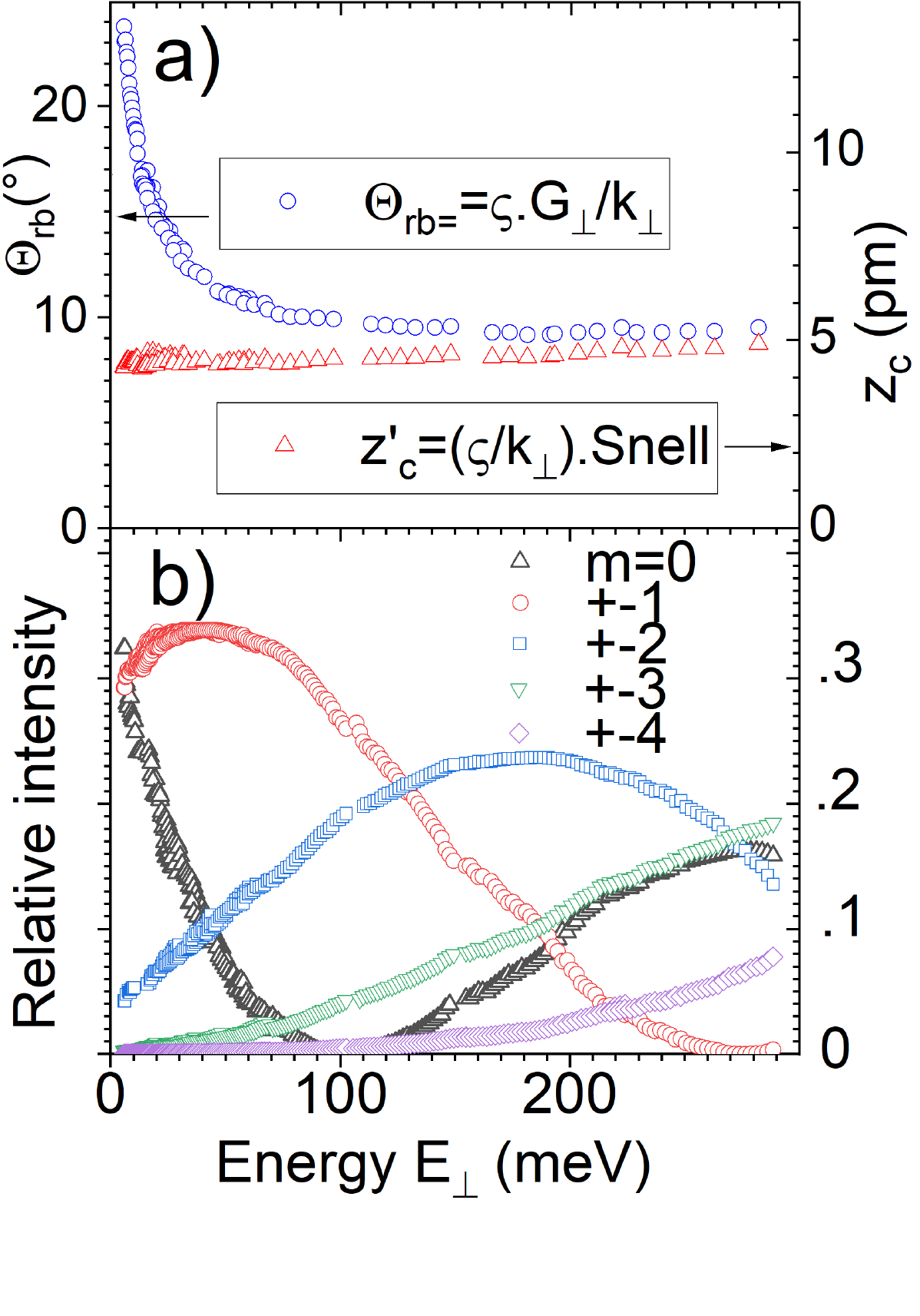}
    \caption{b) Diffracted intensities $I_{\pm m}$=($I_{-m}$ + $I_m$)/2 for 500 eV Ar atoms along the [100] direction. Below 50 meV all intensities $I_m$ were left free but above, a fit forcing a Bessel form $I_m=J_m^2(\zeta)$ was used. a) The derived rainbow angle (\myholowcircle{blue}) $\Theta_{rb}\simeq \zeta\frac{G_\perp }{k_\perp}=\zeta\frac{\lambda_\perp }{a_\perp}$ on the lhs is a good description of the actual value while, interpreting $\zeta$ as a corrugation amplitude $z_c=\zeta/k_\perp$ on the rhs shows an unphysical increase at low energy. Considering refraction with Eq.~\ref{eq:Snell} and $D=50$ meV provides a more plausible value of the actual corrugation (\myholowtriangle{red}). }
    \label{fgr:Ar[100]_Im&zc}
\end{figure}

\begin{figure}[h]
\includegraphics[width=0.85\linewidth,trim={0 0 0 0},clip,angle =0,draft = false]{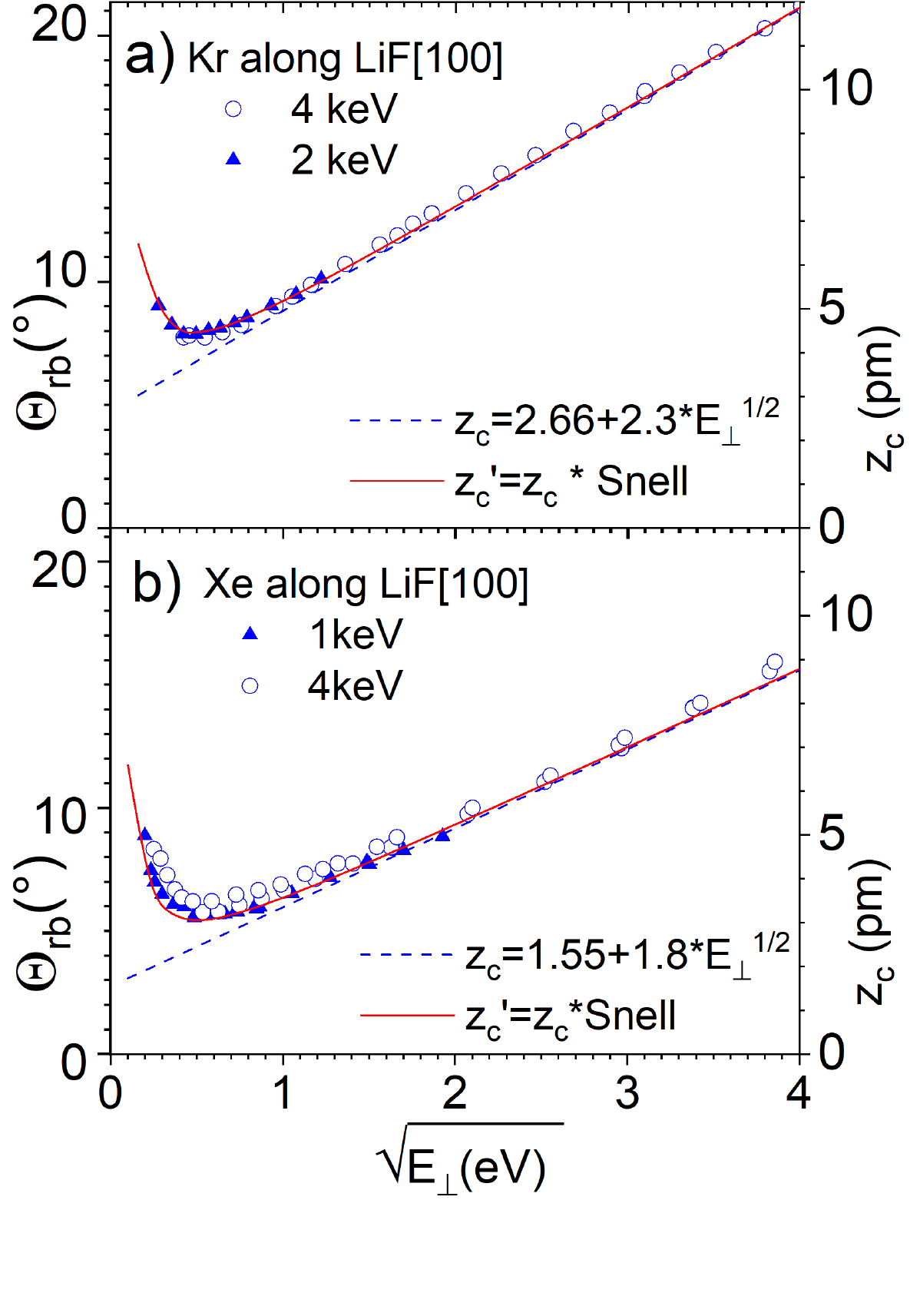}%
\caption{Rainbow scattering angle of Kr and Xe atoms along the [100] direction of LiF[100]. The rainbow angle is tracked by fitting the scattering profile with Bessel functions as in Fig.~\ref{fgr:Kr_Xe_Fit_HW}. The straight dashed lines have been adjusted by hand so that the red line corrected by refraction in Eq.~\ref{eq:Snell}, pass through the data.
The well-depth  used are $D=92$ and $D=135$ meV for Kr and Xe. Note that $E_\perp$ goes up to 16 eV.}\label{fgr:Kr_zc_de_Ep}
\end{figure}

Figs.~\ref{fgr:Ar[100]_Im&zc},~\ref{fgr:Kr_zc_de_Ep}a) and~\ref{fgr:Kr_zc_de_Ep}b) show the measured rainbow angle for Ar, Kr and Xe respectively. 
Fig.~\ref{fgr:Ar[100]_Im&zc} also displays the derived diffracted intensities $I_m$ that can be compared with those of~helium~\cite{Winter_PSS_2011,zugarramurdi2013surface} and~neon~\cite{debiossac2023elastic}.

\section{Larger impact energy, Classical profile}\label{ch:classical}

We use the same HCW model in spite of the fact that all the trajectories should add up incoherently (without attached phase). This is not a problem since using the experimental line width of a few Bragg angles washes out all interferences converging to the classical envelope. In addition, the optical treatment of the slowly varying phase properly deals with the complex problem of rainbow divergence in classical scattering.

\begin{figure}
    \centering
    \includegraphics[width=1\linewidth,trim={0cm 0 14cm 7cm},clip]{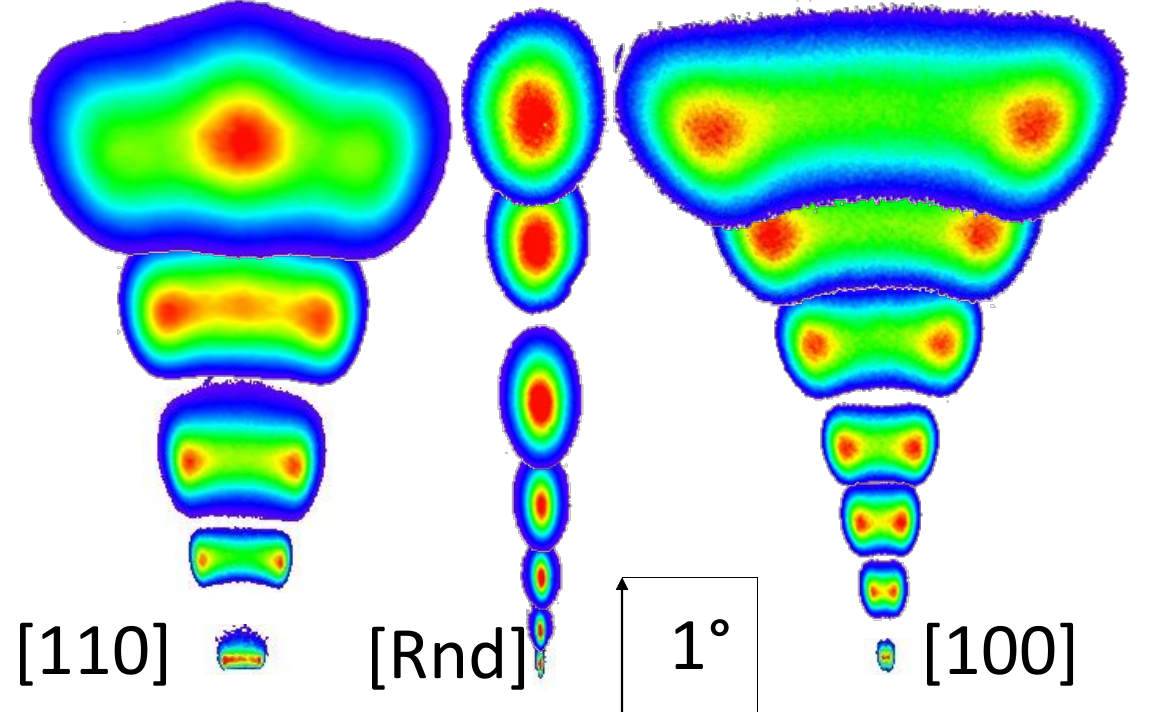}
    \caption{Raw 2D ($\theta_f,\phi_f$) scattering profiles of 4 keV Xe atoms ($\lambda_{//}\sim 40$ fm) along [Rnd], [110], and [100] where $\phi_B$ is 0.008$^\circ$ and 0.011$^\circ$, respectively. The maximum values of $\theta_i$ is around 4.3$^\circ$ corresponding to $E_\perp\sim$ 23 eV. 
    The square indicates the angular scale with the vertical arrow on its side referred to the surface plane $\theta_f=0^\circ$.}
    \label{fig:Xe_4k_2D}
\end{figure}

Figure~\ref{fig:Xe_4k_2D} displays the evolution of the raw 2D scattering profiles recorded for 4 keV Xe atoms along the [110], [Rnd] and [100] directions between a few tens of meV and a few tens of eV. Apart from the lowest energies hardly visible in the bottom but already plotted in Fig.~\ref{fig:Xe_[110]} and Fig.~\ref{fgr:Kr_Xe_Fit_HW}b), most of the scattering profiles have no or very weak residual quantum structures but they still have distinct features such as the central spot on the top left side. 
The latter is characteristic of the [110] direction and is observed for all noble gases appearing at increasing values of $E_\perp$ along the He-Xe sequence. 
For He it is visible above 2 eV, see Fig.~2-3 of Ref.~\cite{Winter_PSS_2011} and Fig. 18 of Ref.~\cite{debiossac2021grazing}, while it appears here for Xe for $E_\perp\gtrsim$ 10 eV. 
It corresponds to a second rainbow peak which is neither a secondary nor a supernumerary rainbow and will be discussed in the next section.
Figure~\ref{fig:Xe_4k_2D} shows that at large values of $E_\perp$ the rainbow angle associated with the [100] direction is significantly larger than that of the [110] direction. This is in contrast with the low energy range, $E_\perp<$ 0.5 eV, where the rainbow angle along [110] is close to 30$^\circ$ while it is closer to 5-8$^\circ$ along [100] for all noble gases.
Note also that the sharp increase of the relative width $\sigma_\Theta \sim \sigma_\phi/\theta_i$ visible in Fig.~\ref{fgr:Kr_zc_de_Ep}b) is hardly seen in the absolute scale of Fig.~\ref{fig:Xe_4k_2D} where $\sigma_\phi$ is visible.
Fig.~\ref{fig:Xe_4k_2D} also shows that the line width defined by the scattering profile recorded along the [Rnd] direction becomes much larger than the Bragg angle $\phi_B$ and finally compares with the width of the rainbow peak.

\section{data analysis and modeling}\label{ch:analysis}
Before discussing the minor differences along the He-Xe sequence, we analyze the common feature appearing in the high energy range where the atomic arrangement along [100] or [110] are decisive and, in the low energy range where attractive force play the dominant role.
Since no specific quantum effect has been observed we use the simple HCW model already mentioned in the data reduction procedure.

The hard corrugated wall model considers free propagation of the atom and instant reflection on a corrugated wall. 
In terms of physics it acknowledges that, due to the exponential character of the Pauli repulsion, most of the momentum transfer is indeed located close to the equipotential surface of energy $E_\perp$. 
As an optical model it takes into account interferences between all rays (trajectories). 
This does not mean that the dynamic of the atom is a good approximation, it simply means that above the surface, most of the phase differences arise from the path difference due to the topology described by the corrugation function. 
This was illustrated recently~\cite{Pollak2024GIFAD} using first order perturbation theory with a realistic Morse potential along $z$ and shifted along $y$ according to a cosine shape. 
The model has the full dynamic of the incoming trajectories inside the Morse potential but
the diffracted intensities are still given by $J_m^2(k_\perp z_c)$. This is because the smooth acceleration and more sudden deceleration when approaching the iso-energy curve is the same for all incoming trajectories leaving the path difference due to the topology $z(y)$ as the only contribution. 
The correction due to the outgoing trajectories appears with second order perturbation theory ~\cite{pollak2015second} and placing the potential well at different location~\cite{allison2022perturbation} also breaks the direct correspondence with the HCW~\cite{Off_axis}.

\subsection{High energy part, detailed topology}
The higher is the energy, the closer we approach the surface and the larger the contrast because the region between surface atoms offers less resistance. 
This describes well the [100] direction where no surface atom lies between the atomics rows.

For He and Ne, it was already noted that the evolution of the corrugation amplitude is quite different along the [100] and [110] directions \cite{momeni2010grazing,debiossac2023elastic}.
The simple interpretation detailed is that along the [100] direction the surface appears as made of a single type of atomic row (see inset in Fig.~\ref{fgr:PEL_corrugation_All}). The maximum of the PEL sits on top of such a row where F$^-$ and Li$^+$ ions alternate while the minimum lies in between. 
At higher energy, the minimum decreases faster (is softer) than the maximum resulting in a corrugation amplitude that increases with energy as visible above $E_\perp=0.5$ eV on the rhs scale of Fig.~\ref{fgr:Kr_zc_de_Ep} where the value $\zeta$ derived from the HCW is converted into a corrugation amplitude.
Along the [110] direction the lowest point in the PEL corresponds to a row of Li$^+$ ions, these have a reduced electron density but their K shell electrons are more deeply bound and more difficult to penetrate while the larger electron density of the F$^-$ row is comparatively softer so that the corrugation amplitude is more or less constant at low energy (see \textit{e.g.} Fig. 3-10 and 3-3 of Ref.~\cite{Winter_PSS_2011} resp.) with a tendency to decrease at the energy $E_\perp$ above a few eV and in Fig.~\ref{fig:Xe_4k_2D} for Xe.

This can be analyzed in more detail by investigating the exact shape of the PEL but the most convincing evidence of the presence of the Li$^+$ row is the bright spot appearing on the top left of Fig.~\ref{fig:Xe_4k_2D}. 
As mentioned above the same spot have been observed for He, Ne, Ar and Kr at increasing values of $E_\perp$ and it has a simple classical origin in terms of a second rainbow structure. 
Assuming that the energy $E_\perp$ is large enough to reveal directly the presence of the Li$^+$ row in the center of the lattice cell, \textit{i.e.} that the iso-energy curve shows two maxima, a big one on the F$^-$ row and a smaller one on the Li$^+$ row, the deflection function will have two extrema associated with the two inflection point in the PEL and two new rainbow peak at positive and negative values of the associated scattering angle. 
Before reaching this situation, the bottom of the PEL flattens and a peak appears in the scattering distribution and its intensity increases progressively. 

For He at $E_\perp$=4.2 eV, this was modeled in Fig.18 of Ref.~\cite{debiossac2021grazing} using the HCW model with a corrugation function having two Fourier components $Z(y)=h_1\cos\tilde{y}+h_2\cos2\tilde{y}$ with $\tilde{y}=G_\perp y$ and $h_2\sim h_1/10$. 
For the pattern in Fig.~\ref{fig:Xe_4k_2D} with Xe at $E_\perp\sim$ 23 eV, the HCW using $h_1=18.8$ pm and $h_2=3.6$ pm produces a good fit made of $\sim$ 250 diffraction orders which are convoluted by a broad gaussian profile with $\sigma_{lw}\simeq$ 30 $\phi_B$. 
The width $\sigma_{lw}$ is so large that the same result can certainly be derived from classical models without any interference but the HCW formula (Eq.2 in Ref.~\cite{debiossac2021grazing}) is so simple that the calculation is instantaneous and cover the full semi-classical range merging smoothly to classical behavior. 

In practice tough, for the values of $h_1$ and $h_2$ used the iso-energy curve $Z(y)$ does not yet have a double maxima shape, however the bottom of the corrugation function $Z(y)$ becomes so flat that the quasi specular reflection in this region produces the intense maximum observed in the center of the scattering profile. 
Note that the curvature $\ddot{Z}(y)\propto-h_1\cos\tilde{y}-4\,h_2\cos2\tilde{y}$ already shows a marked double well structure but the point representing the Li$^+$ rows does not reach zero ($h_2 < h_1/4$) so that there is no inversion of the curvature. 
In other words, the observed central peak is a precursor of the genuine second classical rainbow peak \cite{miret2012classical}.

For He where the same situation takes place in a semi-quantum regime, supernumerary rainbows are also visible on Fig. 16 of Ref.~\cite{debiossac2021grazing}, probably allowing more details. 
With He along KCl[110], the ratio between ionic radii and lattice parameter is such that the same transition takes place close to the quantum regime, between 30 meV and 500 meV producing an apparent disorder in the diffraction charts reporting the evolution of the diffracted intensities \cite{Meyer_thesis,bocan2018gifad,del2020accurate} before a clear second "inner" rainbow structure becomes visible at $E_\perp$=0.9 eV \cite{specht2011rainbow}.

\subsection{Low energy part, attractive forces}

\begin{table}  
    \centering
    \begin{tabular}{c||c||c|c||c|}
          & $D$(meV) & $\Theta_{0 [100]}$($^\circ$) & $\alpha_{[100]}$ ($^\circ /\sqrt{eV}$)& $\Theta_{0 [110]}$($^\circ$) \\

        \hline
        He &  -8.5 & 8.0 & 16& 36 \\
        \hline
        Ne &  -10 & 9.8 & 8.9& 28\\
        \hline
        Ar &  -50 & 6.24 & 5.2 & 23 \\
        \hline
        Kr & -92 & 4.7 & 4.1& 19 \\
        \hline
        Xe & -135 & 2.76 & 3.2& 13 \\
        \hline
    \end{tabular}
        \caption{Parameters used to model our data by the refraction effect in Eq.~\ref{eq:Snell}$: \Theta_{surf}=\Theta_{0}+\alpha \sqrt{E_\perp}$. The energy range is limited along [110] and $\alpha_{[110]}$ was taken as 0 while the broader range investigated along [100] forces the use of an energy dependence. The corresponding corrugation amplitude is $z_c=G_\perp\Theta_{surf}$.}
    \label{tab:phi_0}
\end{table}

For all projectiles and for both [100] and [110] directions, all data indicate a rapid increase of the rainbow angle $\Theta_{rb}$ when the effective energy $E_\perp$ approaches zero. 
The effect is known for long and, within the HCW, was modeled as the Beeby correction~\cite{Beeby1971}. 
let's assume that the attractive forces are important at comparatively large distance from the surface. By replacing the initial wave vector $k_\perp$, associated with $E_\perp$, by $k_\perp'$ ,associated with $E_\perp'=E_\perp + D$, takes into account refraction of the rainbow angle following Snell's law. This also modifies the velocity $\propto k_\perp'$ and the phase shift $k'z_c$ associated with a given path difference $z_c$ (see \textit{e.g.} the review \cite{Farias1998}).

Along the [110] direction we have focused on the rainbow angle position which is large enough to be pointed by hand and we have adjusted the value $\Theta_{surf}$ by hand so that the refracted curve passes through the data in Fig.\ref{fgr:Ar_1784_110_Rb_peaks} and Fig.\ref{Kr_110_4k_rainbow} for Ar and Kr respectively. The data for Xe have not been plotted as no supernumerary rainbow was clearly identified but the same procedure was applied.

Along the [100] the rainbow angle is comparatively small and we have used the very good agreement with the Bessel function $I_m=J_m^2(\zeta)$ to derived precise values of the rainbow angle $\Theta_{rb}\simeq\zeta\frac{\lambda_\perp }{a_\perp}$ describing accurately the observed scattering distributions. 
For Ar, to recover the emission angle $\Theta_{surf}$, we have applied the correction directly to the measured values in Fig.~\ref{fgr:Ar_1784_110_Rb_peaks} and extracted the value $\Theta_{surf}$. For Kr and Xe where the energy range is larger, we have adjusted the linear form $\Theta_{surf}=\Theta_{0}+\alpha\sqrt{E_\perp}$ by hand so that the refracted curve passes through the data in Fig.~\ref{Kr_110_4k_rainbow}a and Fig.~\ref{Kr_110_4k_rainbow}b respectively. 
All the values derived along both [100] and [110] directions are reported in Tab.~\ref{tab:phi_0} using published values \cite{hoinkes1980physical} of the well-depth $D$ .

\subsection{Model derivation of the PEL along [100]}

The previous paragraphs have shown that the Beeby correction does significantly modify the apparent corrugation amplitude derived from the HCW. We have also shown that the same well-depth $D$ of the mean planar potential can be applied along both directions\cite{same_D} to produce a more realistic energy behavior. 
However, this latter was in part forced to follow a given arbitrary analytic form.
This can also be related to an intrinsic weakness of the HCW, each measurement at an energy $E_\perp$ produces a value of $\zeta$ without any connection between them, \textit{i.e.} ignoring the consistency between the iso-energy curves originating essentially from the quasi-exponential decay range of the PEL along the $z$ direction. 

A better consistency can be achieved by fitting all the data simultaneously with a model PEL so that the corrugation function $z_E(y,z)$ are forced to be the iso-energy curves. 
Following Ref.~\cite{debiossac2020refraction}, we use a PEL based on binary potentials developed on a set of screened coulombic forms $V(r)=\Sigma_i(A_i/r) e^{-B_ir}$ attached to atomic positions. 
Such forms are widely used in atomic collision to describe the repulsive part~\cite{zbl}. 
For the attractive part, screened coulomb forms are also well-suited to describe the projectile polarization by the exponentially decaying Madelung field above the surface while van der Walls or Casimir-Polder terms are better modeled by Lennard-Jones term $-B/r^6$ summing up to planar $-B/4z^4$ dependence ending as $\propto z^{-3}$ when retardation effects are taken into account~\cite{lecoffre2025measurement}. 
For neon projectiles, the inclusion of the $\propto z^{-3}$ parameter does affect the screened-coulombic attractive term but did not improve the quality of the fit~\cite{debiossac2020refraction} to the data which appeared to be sensitive mainly to the resulting well-depth $D$. We thus discard this planar term to keep the number of adjustable parameters as low as possible. 

We will compare only the scattering data along the [100] direction where the data reduction to a single number $\zeta(E_\perp)$ was established in Sec.\ref{ch:inelastic} allowing a straightforward search of the topological corrugation function by only two point, one on top of the atomic rows and one in between while the well-Depth $D$ used for refraction is taken from the planar form of the PEL.
The resulting PEL integrated along [100] should be qualitatively correct and the parameters could be considered as a front-end before quantum refinement. 

\begin{figure}
\includegraphics[width=0.9\linewidth,trim={0 0 0 0cm},clip,angle =0,draft = false]{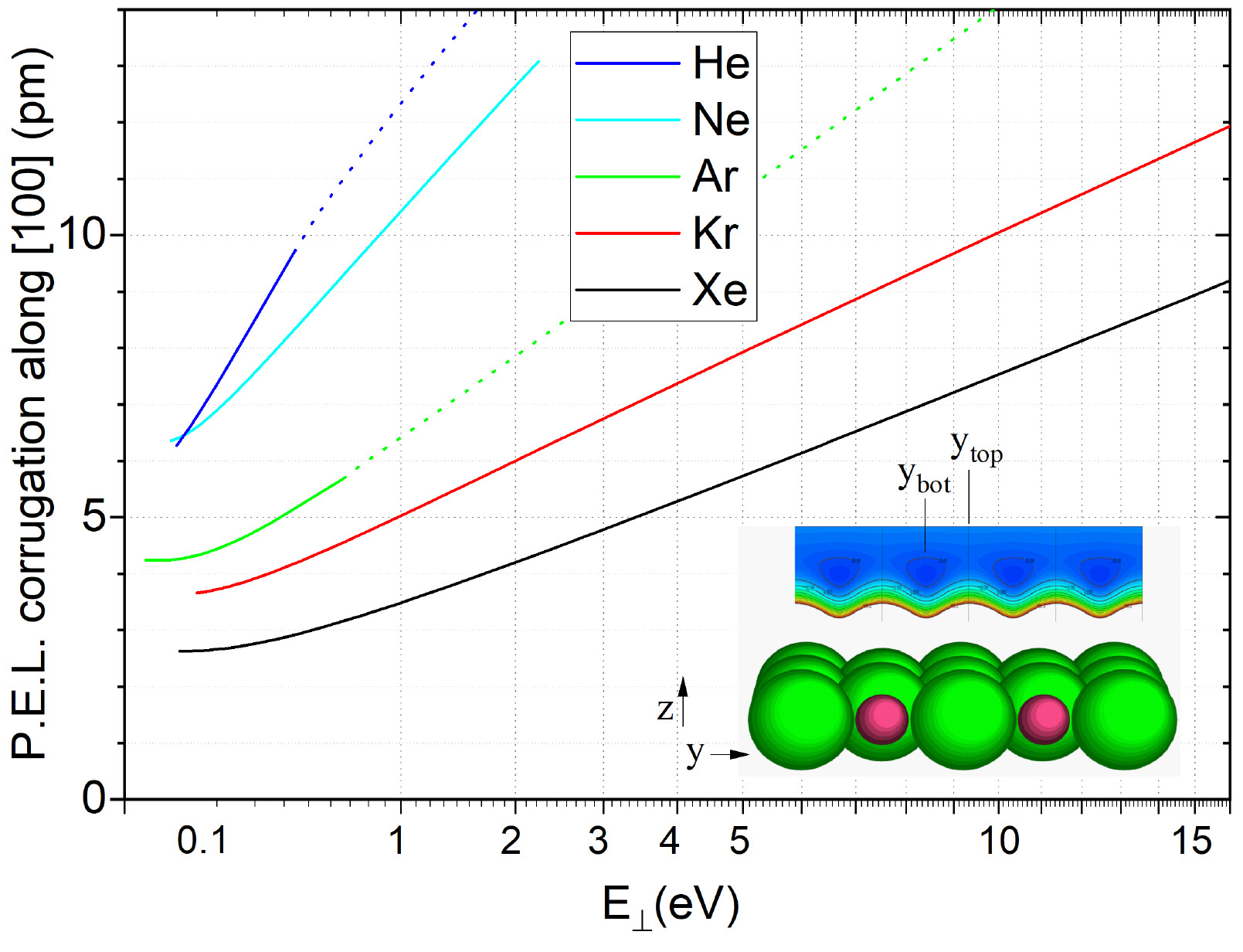}
\caption{\label{fgr:PEL_corrugation_All} Real space corrugation $z_c=z_{E}(y_{top})-z_{E}(y_{bot})$ of the average [100] potential energy landscape as a function of $E_\perp$ plotted in a $\sqrt{E}$ scale. $z_{E}(y)$ are the equipotential lines of the PEL fitted to the diffraction data using the Hardwall model with angular refraction (Beeby correction).
The dotted lines correspond to energy domains where no experimental data were available. 
}
\end{figure}

We use two screened coulomb terms, one for the repulsive part and one for the attractive so that only four terms are needed to build the PEL and predict the diffracted intensities $I_m$ at all energies. 
In practice we have fitted the measured $\zeta$ by $k_\perp'z_{PEL}$ where $z_{PEL}(E_\perp)$ is the actual top to bottom corrugation of the $E=E_\perp$ iso-energy curve  and $k_\perp'$ is the Beeby corrected wave-vector taking refraction into account ($E_\perp'=E_\perp+D$). 
The corrugation amplitude fitted to the data are plotted in Fig.~\ref{fgr:PEL_corrugation_All} for all the noble gases. 
For $E_\perp$ above one eV, the values are not very different from the apparent corrugation reported in previous section by assigning $z_c=\zeta/k_\perp$ but they do differ significantly below and in a way which could not be model by straight line interpolations as in Figs \ref{fgr:Ar_1784_110_Rb_peaks}, \ref{Kr_110_4k_rainbow}, \ref{fgr:Ar[100]_Im&zc} and \ref{fgr:Kr_zc_de_Ep}. 
Fig.~\ref{fgr:PEL_corrugation_All} indicates a distinct leveling of the corrugation amplitude at low energy, even when the contribution of the attractive forces on the projectile trajectories have been compensated. 
This leveling are here properties of the PEL where the attractive forces have compressed all positive iso-energy line below the one corresponding to $E_\perp$=0 rather than letting these lines expand smoothly to the vacuum.

It should be stressed that the fit is comparatively robust, not only because of the redundancy provided by the reduced number of parameters but also because the fitted parameter $\zeta$ is a monotonous function where no oscillation is present avoiding the difficult situation where the fit is trapped in a local minima.

\subsection{perspectives}

Extending our model to fit simultaneously the [110] data requires abandoning $\zeta$ which offers a reliable description of the data only along the [100] direction. 
The fitting parameter could be $\sigma_m$ (or $\sigma_\phi$) which is equivalent to $\zeta$ along [100] but is also perfectly defined along [110] and which can be extracted directly from the raw scattering profile. 
The data reduction procedure used above to extract the diffracted intensities $I_m$ is a kind of de-convolution which becomes more and more difficult when the inelastic line width exceeds the Bragg angle. 
It could be replaced by a convolution of the predicted intensities by the line profile recorded along the [Rnd] direction. 
Such a global fit of all directions would be even more robust and precise as it would be constrained by the supernumerary rainbow structure and could reveal the contribution of the Li$^+$ ions. 
The main perspective is however to use the derived parameters of the binary interaction potential as a seed for a fit by a quantum scattering code to produce values as exact as possible.

\section{conclusion}\label{ch:conclusion}

In GIFAD the role of attractive forces was first overlooked~\cite{Winter_PSS_2011}, in part because with He at low $E_\perp$, only a few diffraction orders are populated and the rainbow angle is not readily visible. 
The corresponding shift in the energy scale was first identified when comparing He diffraction on the $\beta_2 (2\times 4)$ reconstruction of the GaAs(001) surface with \textit{ab initio} calculations~\cite{Debiossac_PRB_2014}.  The Beeby correction was used in GIFAD for He scattered from Graphene/SiC where the corrugation function was describing both the interatomic potential and the shape of the Moiré pattern~\cite{Debiossac_PRB_2016}. 
Then the refraction effect was made model-free in Fig.21 of Ref.~\cite{debiossac2021grazing} by using the standard deviation $\sigma_m$ or $\sigma_\Theta$, bypassing the concept of rainbow angle.
With neon on LiF[110] the reduced wavelength increases the number of diffraction orders making the refraction effect readily visible in the evolution of scattering profiles (Fig. 2 of Ref.~\cite{debiossac2020refraction}). 
By comparing with a quantum scattering code, the Snell's correction was shown to be a decent description of the refraction effect. 
In this paper, the refraction Ar, Kr and Xe is shown to be quite large both along the [110] and [100] directions and could be described consistently with a single value of the well-depth, whereas some publications have suggested that Van der Walls effect are important only along the [110] direction.


The main interest of this semi-classical approach is that it is conceptually and numerically simple and the fit to all data takes place in a few second on an office PC. 
It could, for instance be used as a front-end to initiate the fit by quantum scattering codes offering both accuracy and exactitude. 

In the case of a cosine corrugation function where the diffracted intensities are well-fitted by Bessel function $I_m=J_m^2(\zeta)$, the parameter $\zeta$ is most often interpreted in terms of surface corrugation $z_c=\zeta/k_\perp$ while it describes accurately the rainbow angle $\Theta_{rb}=\frac{G_\perp \zeta}{k_\perp}$ or the scattering width $\sigma_m$.
The Beeby correction is needed to connect $\zeta$, $\Theta_{rb}$ or $\sigma_m$ to the topological corrugation amplitude $z_c$ taking into account the well-depth $D$ of the attractive well.

\section{Acknowledgment}
We are grateful to Hynd Remita for the irradiation of the LiF samples by $\gamma$ rays from the Cobalt source of the Institut de Chimie Physique at Orsay,  favoring cleaving with large terraces. 
We are indebted to J.R. Manson for teaching us part of his background on atomic diffraction a low energy. 
We also acknowledge the enthusiastic participation of Alex Le Guen and Jaafar Najafi Rad during their internship in the group.
This work received support from LabEx PALM (ANR-10-LABX-0039-PALM) and 
Chinese Scholarship Council (CSC) Grant No. 201806180025.

\bibliography{Neon}

\begin{thebibliography}{63}%
\makeatletter
\providecommand \@ifxundefined [1]{%
 \@ifx{#1\undefined}
}%
\providecommand \@ifnum [1]{%
 \ifnum #1\expandafter \@firstoftwo
 \else \expandafter \@secondoftwo
 \fi
}%
\providecommand \@ifx [1]{%
 \ifx #1\expandafter \@firstoftwo
 \else \expandafter \@secondoftwo
 \fi
}%
\providecommand \natexlab [1]{#1}%
\providecommand \enquote  [1]{``#1''}%
\providecommand \bibnamefont  [1]{#1}%
\providecommand \bibfnamefont [1]{#1}%
\providecommand \citenamefont [1]{#1}%
\providecommand \href@noop [0]{\@secondoftwo}%
\providecommand \href [0]{\begingroup \@sanitize@url \@href}%
\providecommand \@href[1]{\@@startlink{#1}\@@href}%
\providecommand \@@href[1]{\endgroup#1\@@endlink}%
\providecommand \@sanitize@url [0]{\catcode `\\12\catcode `\$12\catcode
  `\&12\catcode `\#12\catcode `\^12\catcode `\_12\catcode `\%12\relax}%
\providecommand \@@startlink[1]{}%
\providecommand \@@endlink[0]{}%
\providecommand \url  [0]{\begingroup\@sanitize@url \@url }%
\providecommand \@url [1]{\endgroup\@href {#1}{\urlprefix }}%
\providecommand \urlprefix  [0]{URL }%
\providecommand \Eprint [0]{\href }%
\providecommand \doibase [0]{https://doi.org/}%
\providecommand \selectlanguage [0]{\@gobble}%
\providecommand \bibinfo  [0]{\@secondoftwo}%
\providecommand \bibfield  [0]{\@secondoftwo}%
\providecommand \translation [1]{[#1]}%
\providecommand \BibitemOpen [0]{}%
\providecommand \bibitemStop [0]{}%
\providecommand \bibitemNoStop [0]{.\EOS\space}%
\providecommand \EOS [0]{\spacefactor3000\relax}%
\providecommand \BibitemShut  [1]{\csname bibitem#1\endcsname}%
\let\auto@bib@innerbib\@empty
\bibitem [{\citenamefont {Zugarramurdi}\ and\ \citenamefont
  {Borisov}(2012)}]{Zugarramurdi_2012}%
  \BibitemOpen
  \bibfield  {author} {\bibinfo {author} {\bibfnamefont {A.}~\bibnamefont
  {Zugarramurdi}}\ and\ \bibinfo {author} {\bibfnamefont {A.~G.}\ \bibnamefont
  {Borisov}},\ }\bibfield  {title} {\bibinfo {title} {Transition from fast to
  slow atom diffraction},\ }\href {https://doi.org/10.1103/PhysRevA.86.062903}
  {\bibfield  {journal} {\bibinfo  {journal} {Phys. Rev. A}\ }\textbf {\bibinfo
  {volume} {86}},\ \bibinfo {pages} {062903} (\bibinfo {year}
  {2012})}\BibitemShut {NoStop}%
\bibitem [{\citenamefont {Pollak}\ \emph {et~al.}(2024)\citenamefont {Pollak},
  \citenamefont {Roncin}, \citenamefont {Allison},\ and\ \citenamefont
  {Miret-Artes}}]{Pollak2024GIFAD}%
  \BibitemOpen
  \bibfield  {author} {\bibinfo {author} {\bibfnamefont {E.}~\bibnamefont
  {Pollak}}, \bibinfo {author} {\bibfnamefont {P.}~\bibnamefont {Roncin}},
  \bibinfo {author} {\bibfnamefont {W.}~\bibnamefont {Allison}},\ and\ \bibinfo
  {author} {\bibfnamefont {S.}~\bibnamefont {Miret-Artes}},\ }\bibfield
  {title} {\bibinfo {title} {Grazing incidence fast atom diffraction: General
  con- siderations, semiclassical perturbation theory and experimental
  implications},\ }\href {https://doi.org/10.1039/d4cp02183e} {\bibfield
  {journal} {\bibinfo  {journal} {Phys. Chem. Chem. Phys.}\ }\textbf {\bibinfo
  {volume} {26}},\ \bibinfo {pages} {25501} (\bibinfo {year}
  {2024})}\BibitemShut {NoStop}%
\bibitem [{\citenamefont {Roncin}\ and\ \citenamefont
  {Debiossac}(2017)}]{Roncin_PRB_2017}%
  \BibitemOpen
  \bibfield  {author} {\bibinfo {author} {\bibfnamefont {P.}~\bibnamefont
  {Roncin}}\ and\ \bibinfo {author} {\bibfnamefont {M.}~\bibnamefont
  {Debiossac}},\ }\bibfield  {title} {\bibinfo {title} {Elastic and inelastic
  diffraction of fast atoms, {Debye-Waller factor, and M}\"ossbauer-lamb-dicke
  regime},\ }\href {https://doi.org/10.1103/PhysRevB.96.035415} {\bibfield
  {journal} {\bibinfo  {journal} {Phys. Rev. B}\ }\textbf {\bibinfo {volume}
  {96}},\ \bibinfo {pages} {035415} (\bibinfo {year} {2017})}\BibitemShut
  {NoStop}%
\bibitem [{\citenamefont {Rousseau}\ \emph {et~al.}(2008)\citenamefont
  {Rousseau}, \citenamefont {Khemliche}, \citenamefont {Bundaleski},
  \citenamefont {Soulisse}, \citenamefont {Momeni},\ and\ \citenamefont
  {Roncin}}]{Rousseau_2008}%
  \BibitemOpen
  \bibfield  {author} {\bibinfo {author} {\bibfnamefont {P.}~\bibnamefont
  {Rousseau}}, \bibinfo {author} {\bibfnamefont {H.}~\bibnamefont {Khemliche}},
  \bibinfo {author} {\bibfnamefont {N.}~\bibnamefont {Bundaleski}}, \bibinfo
  {author} {\bibfnamefont {P.}~\bibnamefont {Soulisse}}, \bibinfo {author}
  {\bibfnamefont {A.}~\bibnamefont {Momeni}},\ and\ \bibinfo {author}
  {\bibfnamefont {P.}~\bibnamefont {Roncin}},\ }\bibfield  {title} {\bibinfo
  {title} {Surface analysis with grazing incidence fast atom diffraction
  ({GIFAD})},\ }\href@noop {} {\bibfield  {journal} {\bibinfo  {journal} {J.
  Phys. Conf. Ser.}\ }\textbf {\bibinfo {volume} {133}},\ \bibinfo {pages}
  {012013} (\bibinfo {year} {2008})}\BibitemShut {NoStop}%
\bibitem [{\citenamefont {Manson}\ \emph {et~al.}(2008)\citenamefont {Manson},
  \citenamefont {Khemliche},\ and\ \citenamefont {Roncin}}]{Manson_PRB_2008}%
  \BibitemOpen
  \bibfield  {author} {\bibinfo {author} {\bibfnamefont {J.~R.}\ \bibnamefont
  {Manson}}, \bibinfo {author} {\bibfnamefont {H.}~\bibnamefont {Khemliche}},\
  and\ \bibinfo {author} {\bibfnamefont {P.}~\bibnamefont {Roncin}},\
  }\bibfield  {title} {\bibinfo {title} {Theory of grazing incidence
  diffraction of fast atoms and molecules from surfaces},\ }\href
  {https://doi.org/10.1103/PhysRevB.78.155408} {\bibfield  {journal} {\bibinfo
  {journal} {Phys. Rev. B}\ }\textbf {\bibinfo {volume} {78}},\ \bibinfo
  {pages} {155408} (\bibinfo {year} {2008})}\BibitemShut {NoStop}%
\bibitem [{\citenamefont {Pan}\ \emph {et~al.}(2021)\citenamefont {Pan},
  \citenamefont {Debiossac},\ and\ \citenamefont {Roncin}}]{pan2021polar}%
  \BibitemOpen
  \bibfield  {author} {\bibinfo {author} {\bibfnamefont {P.}~\bibnamefont
  {Pan}}, \bibinfo {author} {\bibfnamefont {M.}~\bibnamefont {Debiossac}},\
  and\ \bibinfo {author} {\bibfnamefont {P.}~\bibnamefont {Roncin}},\
  }\bibfield  {title} {\bibinfo {title} {Polar inelastic profiles in fast-atom
  diffraction at surfaces},\ }\href
  {https://doi.org/10.1103/PhysRevB.104.165415} {\bibfield  {journal} {\bibinfo
   {journal} {Physical Review B}\ }\textbf {\bibinfo {volume} {104}},\ \bibinfo
  {pages} {165415} (\bibinfo {year} {2021})}\BibitemShut {NoStop}%
\bibitem [{\citenamefont {Rousseau}\ \emph {et~al.}(2007)\citenamefont
  {Rousseau}, \citenamefont {Khemliche}, \citenamefont {Borisov},\ and\
  \citenamefont {Roncin}}]{Rousseau_2007}%
  \BibitemOpen
  \bibfield  {author} {\bibinfo {author} {\bibfnamefont {P.}~\bibnamefont
  {Rousseau}}, \bibinfo {author} {\bibfnamefont {H.}~\bibnamefont {Khemliche}},
  \bibinfo {author} {\bibfnamefont {A.~G.}\ \bibnamefont {Borisov}},\ and\
  \bibinfo {author} {\bibfnamefont {P.}~\bibnamefont {Roncin}},\ }\bibfield
  {title} {\bibinfo {title} {Quantum scattering of fast atoms and molecules on
  surfaces},\ }\href {https://doi.org/10.1103/PhysRevLett.98.016104} {\bibfield
   {journal} {\bibinfo  {journal} {Phys. Rev. Lett.}\ }\textbf {\bibinfo
  {volume} {98}},\ \bibinfo {pages} {016104} (\bibinfo {year}
  {2007})}\BibitemShut {NoStop}%
\bibitem [{\citenamefont {Sch\"uller}\ \emph {et~al.}(2010)\citenamefont
  {Sch\"uller}, \citenamefont {Wethekam}, \citenamefont {Blauth}, \citenamefont
  {Winter}, \citenamefont {Aigner}, \citenamefont
  {Simonovi\ifmmode~\acute{c}\else \'{c}\fi{}}, \citenamefont {Solleder},
  \citenamefont {Burgd\"orfer},\ and\ \citenamefont {Wirtz}}]{Schuller2010}%
  \BibitemOpen
  \bibfield  {author} {\bibinfo {author} {\bibfnamefont {A.}~\bibnamefont
  {Sch\"uller}}, \bibinfo {author} {\bibfnamefont {S.}~\bibnamefont
  {Wethekam}}, \bibinfo {author} {\bibfnamefont {D.}~\bibnamefont {Blauth}},
  \bibinfo {author} {\bibfnamefont {H.}~\bibnamefont {Winter}}, \bibinfo
  {author} {\bibfnamefont {F.}~\bibnamefont {Aigner}}, \bibinfo {author}
  {\bibfnamefont {N.}~\bibnamefont {Simonovi\ifmmode~\acute{c}\else
  \'{c}\fi{}}}, \bibinfo {author} {\bibfnamefont {B.}~\bibnamefont {Solleder}},
  \bibinfo {author} {\bibfnamefont {J.}~\bibnamefont {Burgd\"orfer}},\ and\
  \bibinfo {author} {\bibfnamefont {L.}~\bibnamefont {Wirtz}},\ }\bibfield
  {title} {\bibinfo {title} {Rumpling of {LiF}(001) surface from fast atom
  diffraction},\ }\href {https://doi.org/10.1103/PhysRevA.82.062902} {\bibfield
   {journal} {\bibinfo  {journal} {Phys. Rev. A}\ }\textbf {\bibinfo {volume}
  {82}},\ \bibinfo {pages} {062902} (\bibinfo {year} {2010})}\BibitemShut
  {NoStop}%
\bibitem [{\citenamefont {Muzas}\ \emph {et~al.}(2016)\citenamefont {Muzas},
  \citenamefont {Gatti}, \citenamefont {Martín},\ and\ \citenamefont
  {Díaz}}]{Diaz_2016b}%
  \BibitemOpen
  \bibfield  {author} {\bibinfo {author} {\bibfnamefont {A.}~\bibnamefont
  {Muzas}}, \bibinfo {author} {\bibfnamefont {F.}~\bibnamefont {Gatti}},
  \bibinfo {author} {\bibfnamefont {F.}~\bibnamefont {Martín}},\ and\ \bibinfo
  {author} {\bibfnamefont {C.}~\bibnamefont {Díaz}},\ }\bibfield  {title}
  {\bibinfo {title} {Diffraction of {H from LiF(001)}: From slow normal
  incidence to fast grazing incidence},\ }\href
  {https://doi.org/https://doi.org/10.1016/j.nimb.2016.04.031} {\bibfield
  {journal} {\bibinfo  {journal} {NIM--B}\ }\textbf {\bibinfo {volume} {382}},\
  \bibinfo {pages} {49 } (\bibinfo {year} {2016})}\BibitemShut {NoStop}%
\bibitem [{\citenamefont {Zugarramurdi}\ \emph {et~al.}(2013)\citenamefont
  {Zugarramurdi}, \citenamefont {Debiossac}, \citenamefont {Lunca-Popa},
  \citenamefont {Alarc{\'o}n}, \citenamefont {Momeni}, \citenamefont
  {Khemliche}, \citenamefont {Roncin},\ and\ \citenamefont
  {Borisov}}]{zugarramurdi2013surface}%
  \BibitemOpen
  \bibfield  {author} {\bibinfo {author} {\bibfnamefont {A.}~\bibnamefont
  {Zugarramurdi}}, \bibinfo {author} {\bibfnamefont {M.}~\bibnamefont
  {Debiossac}}, \bibinfo {author} {\bibfnamefont {P.}~\bibnamefont
  {Lunca-Popa}}, \bibinfo {author} {\bibfnamefont {L.}~\bibnamefont
  {Alarc{\'o}n}}, \bibinfo {author} {\bibfnamefont {A.}~\bibnamefont {Momeni}},
  \bibinfo {author} {\bibfnamefont {H.}~\bibnamefont {Khemliche}}, \bibinfo
  {author} {\bibfnamefont {P.}~\bibnamefont {Roncin}},\ and\ \bibinfo {author}
  {\bibfnamefont {A.}~\bibnamefont {Borisov}},\ }\bibfield  {title} {\bibinfo
  {title} {Surface-grating deflection of fast atom beams},\ }\href
  {https://doi.org/10.1103/PhysRevA.88.012904} {\bibfield  {journal} {\bibinfo
  {journal} {Physical Review A.}\ }\textbf {\bibinfo {volume} {88}},\ \bibinfo
  {pages} {012904} (\bibinfo {year} {2013})}\BibitemShut {NoStop}%
\bibitem [{\citenamefont {Jardine}\ \emph {et~al.}(2004)\citenamefont
  {Jardine}, \citenamefont {Dworski}, \citenamefont {Fouquet}, \citenamefont
  {Alexandrowicz}, \citenamefont {Riley}, \citenamefont {Lee}, \citenamefont
  {Ellis},\ and\ \citenamefont {Allison}}]{jardine2004ultrahigh}%
  \BibitemOpen
  \bibfield  {author} {\bibinfo {author} {\bibfnamefont {A.~P.}\ \bibnamefont
  {Jardine}}, \bibinfo {author} {\bibfnamefont {S.}~\bibnamefont {Dworski}},
  \bibinfo {author} {\bibfnamefont {P.}~\bibnamefont {Fouquet}}, \bibinfo
  {author} {\bibfnamefont {G.}~\bibnamefont {Alexandrowicz}}, \bibinfo {author}
  {\bibfnamefont {D.~J.}\ \bibnamefont {Riley}}, \bibinfo {author}
  {\bibfnamefont {G.~Y.}\ \bibnamefont {Lee}}, \bibinfo {author} {\bibfnamefont
  {J.}~\bibnamefont {Ellis}},\ and\ \bibinfo {author} {\bibfnamefont
  {W.}~\bibnamefont {Allison}},\ }\bibfield  {title} {\bibinfo {title}
  {Ultrahigh-resolution spin-echo measurement of surface potential energy
  landscapes},\ }\href {https://doi.org/10.1126/science.1098490} {\bibfield
  {journal} {\bibinfo  {journal} {Science}\ }\textbf {\bibinfo {volume}
  {304}},\ \bibinfo {pages} {1790} (\bibinfo {year} {2004})}\BibitemShut
  {NoStop}%
\bibitem [{\citenamefont {Debiossac}\ \emph
  {et~al.}(2014{\natexlab{a}})\citenamefont {Debiossac}, \citenamefont
  {Zugarramurdi}, \citenamefont {Lunca-Popa}, \citenamefont {Momeni},
  \citenamefont {Khemliche}, \citenamefont {Borisov},\ and\ \citenamefont
  {Roncin}}]{Debiossac_PRL_2014}%
  \BibitemOpen
  \bibfield  {author} {\bibinfo {author} {\bibfnamefont {M.}~\bibnamefont
  {Debiossac}}, \bibinfo {author} {\bibfnamefont {A.}~\bibnamefont
  {Zugarramurdi}}, \bibinfo {author} {\bibfnamefont {P.}~\bibnamefont
  {Lunca-Popa}}, \bibinfo {author} {\bibfnamefont {A.}~\bibnamefont {Momeni}},
  \bibinfo {author} {\bibfnamefont {H.}~\bibnamefont {Khemliche}}, \bibinfo
  {author} {\bibfnamefont {A.~G.}\ \bibnamefont {Borisov}},\ and\ \bibinfo
  {author} {\bibfnamefont {P.}~\bibnamefont {Roncin}},\ }\bibfield  {title}
  {\bibinfo {title} {Transient quantum trapping of fast atoms at surfaces},\
  }\href {https://doi.org/10.1103/PhysRevLett.112.023203} {\bibfield  {journal}
  {\bibinfo  {journal} {Phys. Rev. Lett.}\ }\textbf {\bibinfo {volume} {112}},\
  \bibinfo {pages} {023203} (\bibinfo {year} {2014}{\natexlab{a}})}\BibitemShut
  {NoStop}%
\bibitem [{\citenamefont {Zhao}\ \emph {et~al.}(2008)\citenamefont {Zhao},
  \citenamefont {Schulz}, \citenamefont {Meek}, \citenamefont {Meijer},\ and\
  \citenamefont {Sch{\"o}llkopf}}]{zhao2008quantum}%
  \BibitemOpen
  \bibfield  {author} {\bibinfo {author} {\bibfnamefont {B.~S.}\ \bibnamefont
  {Zhao}}, \bibinfo {author} {\bibfnamefont {S.}~\bibnamefont {Schulz}},
  \bibinfo {author} {\bibfnamefont {S.}~\bibnamefont {Meek}}, \bibinfo {author}
  {\bibfnamefont {G.}~\bibnamefont {Meijer}},\ and\ \bibinfo {author}
  {\bibfnamefont {W.}~\bibnamefont {Sch{\"o}llkopf}},\ }\bibfield  {title}
  {\bibinfo {title} {Quantum reflection of helium atom beams from a
  microstructured grating},\ }\href@noop {} {\bibfield  {journal} {\bibinfo
  {journal} {Phys. Rev. A}\ }\textbf {\bibinfo {volume} {78}},\ \bibinfo
  {pages} {010902} (\bibinfo {year} {2008})}\BibitemShut {NoStop}%
\bibitem [{\citenamefont {Holst}\ \emph {et~al.}(2021)\citenamefont {Holst},
  \citenamefont {Alexandrowicz}, \citenamefont {Avidor} \emph
  {et~al.}}]{holst2021material}%
  \BibitemOpen
  \bibfield  {author} {\bibinfo {author} {\bibfnamefont {B.}~\bibnamefont
  {Holst}}, \bibinfo {author} {\bibfnamefont {G.}~\bibnamefont
  {Alexandrowicz}}, \bibinfo {author} {\bibfnamefont {N.}~\bibnamefont
  {Avidor}}, \emph {et~al.},\ }\bibfield  {title} {\bibinfo {title} {Material
  properties particularly suited to be measured with helium scattering:
  selected examples from {2D} materials ...},\ }\href@noop {} {\bibfield
  {journal} {\bibinfo  {journal} {Phys. Chem. Chem. Phys.}\ }\textbf {\bibinfo
  {volume} {23}},\ \bibinfo {pages} {7653} (\bibinfo {year}
  {2021})}\BibitemShut {NoStop}%
\bibitem [{\citenamefont {LeGrand}\ and\ \citenamefont
  {Greene}(1986)}]{legrand1986nearly}%
  \BibitemOpen
  \bibfield  {author} {\bibinfo {author} {\bibfnamefont {A.}~\bibnamefont
  {LeGrand}}\ and\ \bibinfo {author} {\bibfnamefont {E.}~\bibnamefont
  {Greene}},\ }\bibfield  {title} {\bibinfo {title} {The nearly elastic
  scattering of {Ne and Ar from LiF} (100)},\ }\href@noop {} {\bibfield
  {journal} {\bibinfo  {journal} {The Journal of chemical physics}\ }\textbf
  {\bibinfo {volume} {84}},\ \bibinfo {pages} {6483} (\bibinfo {year}
  {1986})}\BibitemShut {NoStop}%
\bibitem [{\citenamefont {Moix}\ and\ \citenamefont
  {Pollak}(2011)}]{moix2011communication}%
  \BibitemOpen
  \bibfield  {author} {\bibinfo {author} {\bibfnamefont {J.~M.}\ \bibnamefont
  {Moix}}\ and\ \bibinfo {author} {\bibfnamefont {E.}~\bibnamefont {Pollak}},\
  }\bibfield  {title} {\bibinfo {title} {Communication: Heavy atom quantum
  diffraction by scattering from surfaces},\ }\href@noop {} {\bibfield
  {journal} {\bibinfo  {journal} {The Journal of chemical physics}\ }\textbf
  {\bibinfo {volume} {134}} (\bibinfo {year} {2011})}\BibitemShut {NoStop}%
\bibitem [{\citenamefont {~}\ \emph {et~al.}(2013)\citenamefont {~},
  \citenamefont {Miraglia}, \citenamefont {Sch{\"u}ller},\ and\ \citenamefont
  {Winter}}]{gravielle2013interaction}%
  \BibitemOpen
  \bibfield  {author} {\bibinfo {author} {\bibfnamefont {M.~S.}\ \bibnamefont
  {~}}, \bibinfo {author} {\bibfnamefont {J.~E.}\ \bibnamefont {Miraglia}},
  \bibinfo {author} {\bibfnamefont {A.}~\bibnamefont {Sch{\"u}ller}},\ and\
  \bibinfo {author} {\bibfnamefont {H.}~\bibnamefont {Winter}},\ }\bibfield
  {title} {\bibinfo {title} {Interaction potentials for multi-electron atoms in
  front of a {LiF} (0 0 1) surface from rainbow scattering},\ }\href
  {https://doi.org/10.1016/j.nimb.2013.02.006} {\bibfield  {journal} {\bibinfo
  {journal} {NIM-B}\ }\textbf {\bibinfo {volume} {317}},\ \bibinfo {pages} {77}
  (\bibinfo {year} {2013})}\BibitemShut {NoStop}%
\bibitem [{\citenamefont {Miraglia}\ and\ \citenamefont
  {Gravielle}(2017)}]{miraglia2017reexamination}%
  \BibitemOpen
  \bibfield  {author} {\bibinfo {author} {\bibfnamefont {J.~E.}\ \bibnamefont
  {Miraglia}}\ and\ \bibinfo {author} {\bibfnamefont {M.~S.}\ \bibnamefont
  {Gravielle}},\ }\bibfield  {title} {\bibinfo {title} {Reexamination of the
  interaction of atoms with a {LiF} (001) surface},\ }\href@noop {} {\bibfield
  {journal} {\bibinfo  {journal} {Physical Review A}\ }\textbf {\bibinfo
  {volume} {95}},\ \bibinfo {pages} {022710} (\bibinfo {year}
  {2017})}\BibitemShut {NoStop}%
\bibitem [{\citenamefont {Mattera}\ \emph {et~al.}(1983)\citenamefont
  {Mattera}, \citenamefont {Rocca}, \citenamefont {Salvo}, \citenamefont
  {Terreni}, \citenamefont {Tommasini},\ and\ \citenamefont
  {Valbusa}}]{mattera1983quasi}%
  \BibitemOpen
  \bibfield  {author} {\bibinfo {author} {\bibfnamefont {L.}~\bibnamefont
  {Mattera}}, \bibinfo {author} {\bibfnamefont {M.}~\bibnamefont {Rocca}},
  \bibinfo {author} {\bibfnamefont {C.}~\bibnamefont {Salvo}}, \bibinfo
  {author} {\bibfnamefont {S.}~\bibnamefont {Terreni}}, \bibinfo {author}
  {\bibfnamefont {F.}~\bibnamefont {Tommasini}},\ and\ \bibinfo {author}
  {\bibfnamefont {U.}~\bibnamefont {Valbusa}},\ }\bibfield  {title} {\bibinfo
  {title} {Quasi-elastic scattering of neon from (001) {LiF} surface},\
  }\href@noop {} {\bibfield  {journal} {\bibinfo  {journal} {Surface Science}\
  }\textbf {\bibinfo {volume} {124}},\ \bibinfo {pages} {571} (\bibinfo {year}
  {1983})}\BibitemShut {NoStop}%
\bibitem [{\citenamefont {Rieder}\ and\ \citenamefont
  {Stocker}(1984)}]{rieder1984observation}%
  \BibitemOpen
  \bibfield  {author} {\bibinfo {author} {\bibfnamefont {K.}~\bibnamefont
  {Rieder}}\ and\ \bibinfo {author} {\bibfnamefont {W.}~\bibnamefont
  {Stocker}},\ }\bibfield  {title} {\bibinfo {title} {Observation of pronounced
  neon diffraction from low-index metal surfaces},\ }\href@noop {} {\bibfield
  {journal} {\bibinfo  {journal} {Physical review letters}\ }\textbf {\bibinfo
  {volume} {52}},\ \bibinfo {pages} {352} (\bibinfo {year} {1984})}\BibitemShut
  {NoStop}%
\bibitem [{\citenamefont {Semerad}\ \emph {et~al.}(1987)\citenamefont
  {Semerad}, \citenamefont {Sequard-Base},\ and\ \citenamefont
  {H{\"o}rl}}]{semerad1987resonant}%
  \BibitemOpen
  \bibfield  {author} {\bibinfo {author} {\bibfnamefont {E.}~\bibnamefont
  {Semerad}}, \bibinfo {author} {\bibfnamefont {P.}~\bibnamefont
  {Sequard-Base}},\ and\ \bibinfo {author} {\bibfnamefont {E.}~\bibnamefont
  {H{\"o}rl}},\ }\bibfield  {title} {\bibinfo {title} {Resonant scatterning of
  {Ne/LiF} (001) measured by time of flight analysis},\ }\href@noop {}
  {\bibfield  {journal} {\bibinfo  {journal} {Surface Science}\ }\textbf
  {\bibinfo {volume} {189}},\ \bibinfo {pages} {975} (\bibinfo {year}
  {1987})}\BibitemShut {NoStop}%
\bibitem [{\citenamefont {Minniti}\ \emph {et~al.}(2012)\citenamefont
  {Minniti}, \citenamefont {D{\'\i}az}, \citenamefont {Cu{\~n}ado},
  \citenamefont {Politano}, \citenamefont {Maccariello}, \citenamefont
  {Mart{\'\i}n}, \citenamefont {Far{\'\i}as},\ and\ \citenamefont
  {Miranda}}]{minniti2012helium}%
  \BibitemOpen
  \bibfield  {author} {\bibinfo {author} {\bibfnamefont {M.}~\bibnamefont
  {Minniti}}, \bibinfo {author} {\bibfnamefont {C.}~\bibnamefont {D{\'\i}az}},
  \bibinfo {author} {\bibfnamefont {J.~F.}\ \bibnamefont {Cu{\~n}ado}},
  \bibinfo {author} {\bibfnamefont {A.}~\bibnamefont {Politano}}, \bibinfo
  {author} {\bibfnamefont {D.}~\bibnamefont {Maccariello}}, \bibinfo {author}
  {\bibfnamefont {F.}~\bibnamefont {Mart{\'\i}n}}, \bibinfo {author}
  {\bibfnamefont {D.}~\bibnamefont {Far{\'\i}as}},\ and\ \bibinfo {author}
  {\bibfnamefont {R.}~\bibnamefont {Miranda}},\ }\bibfield  {title} {\bibinfo
  {title} {Helium, neon and argon diffraction from {Ru} (0001)},\ }\href@noop
  {} {\bibfield  {journal} {\bibinfo  {journal} {Journal of Physics: Condensed
  Matter}\ }\textbf {\bibinfo {volume} {24}},\ \bibinfo {pages} {354002}
  (\bibinfo {year} {2012})}\BibitemShut {NoStop}%
\bibitem [{\citenamefont {Debiossac}\ \emph {et~al.}(2020)\citenamefont
  {Debiossac}, \citenamefont {Roncin},\ and\ \citenamefont
  {Borisov}}]{debiossac2020refraction}%
  \BibitemOpen
  \bibfield  {author} {\bibinfo {author} {\bibfnamefont {M.}~\bibnamefont
  {Debiossac}}, \bibinfo {author} {\bibfnamefont {P.}~\bibnamefont {Roncin}},\
  and\ \bibinfo {author} {\bibfnamefont {A.}~\bibnamefont {Borisov}},\
  }\bibfield  {title} {\bibinfo {title} {Refraction of fast {Ne} atoms in the
  attractive well of a {LiF} (001) surface},\ }\href
  {https://doi.org/10.1021/acs.jpclett.0c01157} {\bibfield  {journal} {\bibinfo
   {journal} {J. Phys. Chem. Lett.}\ }\textbf {\bibinfo {volume} {11}},\
  \bibinfo {pages} {4564} (\bibinfo {year} {2020})}\BibitemShut {NoStop}%
\bibitem [{\citenamefont {Winter}\ and\ \citenamefont
  {Schüller}(2011)}]{Winter_PSS_2011}%
  \BibitemOpen
  \bibfield  {author} {\bibinfo {author} {\bibfnamefont {H.}~\bibnamefont
  {Winter}}\ and\ \bibinfo {author} {\bibfnamefont {A.}~\bibnamefont
  {Schüller}},\ }\bibfield  {title} {\bibinfo {title} {Fast atom diffraction
  during grazing scattering from surfaces},\ }\href
  {https://doi.org/10.1016/j.progsurf.2011.07.001} {\bibfield  {journal}
  {\bibinfo  {journal} {Progress in Surface Science}\ }\textbf {\bibinfo
  {volume} {86}},\ \bibinfo {pages} {169 } (\bibinfo {year}
  {2011})}\BibitemShut {NoStop}%
\bibitem [{\citenamefont {Debiossac}\ \emph {et~al.}(2021)\citenamefont
  {Debiossac}, \citenamefont {Pan},\ and\ \citenamefont
  {Roncin}}]{debiossac2021grazing}%
  \BibitemOpen
  \bibfield  {author} {\bibinfo {author} {\bibfnamefont {M.}~\bibnamefont
  {Debiossac}}, \bibinfo {author} {\bibfnamefont {P.}~\bibnamefont {Pan}},\
  and\ \bibinfo {author} {\bibfnamefont {P.}~\bibnamefont {Roncin}},\
  }\bibfield  {title} {\bibinfo {title} {Grazing incidence fast atom
  diffraction, similarities and differences with thermal energy atom scattering
  {(TEAS)}},\ }\href {https://doi.org/10.1039/D0CP05476C} {\bibfield  {journal}
  {\bibinfo  {journal} {Physical Chemistry Chemical Physics}\ }\textbf
  {\bibinfo {volume} {23}},\ \bibinfo {pages} {7615} (\bibinfo {year}
  {2021})}\BibitemShut {NoStop}%
\bibitem [{\citenamefont {Gravielle}\ \emph {et~al.}(2011)\citenamefont
  {Gravielle}, \citenamefont {Schüller}, \citenamefont {Winter},\ and\
  \citenamefont {Miraglia}}]{Gravielle_2011}%
  \BibitemOpen
  \bibfield  {author} {\bibinfo {author} {\bibfnamefont {M.}~\bibnamefont
  {Gravielle}}, \bibinfo {author} {\bibfnamefont {A.}~\bibnamefont
  {Schüller}}, \bibinfo {author} {\bibfnamefont {H.}~\bibnamefont {Winter}},\
  and\ \bibinfo {author} {\bibfnamefont {J.}~\bibnamefont {Miraglia}},\
  }\bibfield  {title} {\bibinfo {title} {Fast atom diffraction for grazing
  scattering of {Ne} atoms from a {LiF}(001) surface},\ }\href
  {https://doi.org/https://doi.org/10.1016/j.nimb.2010.12.011} {\bibfield
  {journal} {\bibinfo  {journal} {NIM--B}\ }\textbf {\bibinfo {volume} {269}},\
  \bibinfo {pages} {1208 } (\bibinfo {year} {2011})}\BibitemShut {NoStop}%
\bibitem [{\citenamefont {Debiossac}\ \emph {et~al.}(2023)\citenamefont
  {Debiossac}, \citenamefont {Pan},\ and\ \citenamefont
  {Roncin}}]{debiossac2023elastic}%
  \BibitemOpen
  \bibfield  {author} {\bibinfo {author} {\bibfnamefont {M.}~\bibnamefont
  {Debiossac}}, \bibinfo {author} {\bibfnamefont {P.}~\bibnamefont {Pan}},\
  and\ \bibinfo {author} {\bibfnamefont {P.}~\bibnamefont {Roncin}},\
  }\bibfield  {title} {\bibinfo {title} {Elastic and inelastic diffraction of
  fast neon atoms on a {LiF} surface},\ }\href@noop {} {\bibfield  {journal}
  {\bibinfo  {journal} {Physical Chemistry Chemical Physics}\ }\textbf
  {\bibinfo {volume} {25}},\ \bibinfo {pages} {30966} (\bibinfo {year}
  {2023})}\BibitemShut {NoStop}%
\bibitem [{\citenamefont {Pan}\ \emph {et~al.}(2022{\natexlab{a}})\citenamefont
  {Pan}, \citenamefont {Rad},\ and\ \citenamefont {Roncin}}]{pan2022setup}%
  \BibitemOpen
  \bibfield  {author} {\bibinfo {author} {\bibfnamefont {P.}~\bibnamefont
  {Pan}}, \bibinfo {author} {\bibfnamefont {J.~N.}\ \bibnamefont {Rad}},\ and\
  \bibinfo {author} {\bibfnamefont {P.}~\bibnamefont {Roncin}},\ }\bibfield
  {title} {\bibinfo {title} {A setup for grazing incidence fast atom
  diffraction},\ }\bibfield  {journal} {\bibinfo  {journal} {Review of
  Scientific Instruments}\ }\textbf {\bibinfo {volume} {93}},\ \href
  {https://doi.org/10.1063/5.0099269} {10.1063/5.0099269} (\bibinfo {year}
  {2022}{\natexlab{a}})\BibitemShut {NoStop}%
\bibitem [{\citenamefont {Debiossac}\ and\ \citenamefont
  {Roncin}(2016)}]{Debiossac_Nim_2016}%
  \BibitemOpen
  \bibfield  {author} {\bibinfo {author} {\bibfnamefont {M.}~\bibnamefont
  {Debiossac}}\ and\ \bibinfo {author} {\bibfnamefont {P.}~\bibnamefont
  {Roncin}},\ }\bibfield  {title} {\bibinfo {title} {Image processing for
  grazing incidence fast atom diffraction},\ }\href
  {https://doi.org/https://doi.org/10.1016/j.nimb.2016.05.023} {\bibfield
  {journal} {\bibinfo  {journal} {NIM--B}\ }\textbf {\bibinfo {volume} {382}},\
  \bibinfo {pages} {36 } (\bibinfo {year} {2016})}\BibitemShut {NoStop}%
\bibitem [{\citenamefont {Lupone}\ \emph {et~al.}(2018)\citenamefont {Lupone},
  \citenamefont {Soulisse},\ and\ \citenamefont {Roncin}}]{lupone2018large}%
  \BibitemOpen
  \bibfield  {author} {\bibinfo {author} {\bibfnamefont {S.}~\bibnamefont
  {Lupone}}, \bibinfo {author} {\bibfnamefont {P.}~\bibnamefont {Soulisse}},\
  and\ \bibinfo {author} {\bibfnamefont {P.}~\bibnamefont {Roncin}},\
  }\bibfield  {title} {\bibinfo {title} {A large area high resolution imaging
  detector for fast atom diffraction},\ }\href
  {https://doi.org/10.1016/j.nimb.2018.04.030} {\bibfield  {journal} {\bibinfo
  {journal} {NIM-B}\ }\textbf {\bibinfo {volume} {427}},\ \bibinfo {pages} {95}
  (\bibinfo {year} {2018})}\BibitemShut {NoStop}%
\bibitem [{\citenamefont {Pan}\ \emph {et~al.}(2022{\natexlab{b}})\citenamefont
  {Pan}, \citenamefont {Debiossac},\ and\ \citenamefont
  {Roncin}}]{pan2022temperature}%
  \BibitemOpen
  \bibfield  {author} {\bibinfo {author} {\bibfnamefont {P.}~\bibnamefont
  {Pan}}, \bibinfo {author} {\bibfnamefont {M.}~\bibnamefont {Debiossac}},\
  and\ \bibinfo {author} {\bibfnamefont {P.}~\bibnamefont {Roncin}},\
  }\bibfield  {title} {\bibinfo {title} {Temperature dependence in fast-atom
  diffraction at surfaces},\ }\href {https://doi.org/10.1039/d2cp00829g}
  {\bibfield  {journal} {\bibinfo  {journal} {Physical Chemistry Chemical
  Physics}\ }\textbf {\bibinfo {volume} {24}},\ \bibinfo {pages} {12319}
  (\bibinfo {year} {2022}{\natexlab{b}})}\BibitemShut {NoStop}%
\bibitem [{\citenamefont {Seifert}\ \emph {et~al.}(2015)\citenamefont
  {Seifert}, \citenamefont {Lienemann}, \citenamefont {Schüller},\ and\
  \citenamefont {Winter}}]{Seifert_2015}%
  \BibitemOpen
  \bibfield  {author} {\bibinfo {author} {\bibfnamefont {J.}~\bibnamefont
  {Seifert}}, \bibinfo {author} {\bibfnamefont {J.}~\bibnamefont {Lienemann}},
  \bibinfo {author} {\bibfnamefont {A.}~\bibnamefont {Schüller}},\ and\
  \bibinfo {author} {\bibfnamefont {H.}~\bibnamefont {Winter}},\ }\bibfield
  {title} {\bibinfo {title} {Studies on coherence and decoherence in fast atom
  diffraction},\ }\href
  {https://doi.org/https://doi.org/10.1016/j.nimb.2015.01.016} {\bibfield
  {journal} {\bibinfo  {journal} {NIM-B}\ }\textbf {\bibinfo {volume} {350}},\
  \bibinfo {pages} {99 } (\bibinfo {year} {2015})}\BibitemShut {NoStop}%
\bibitem [{\citenamefont {Pan}\ \emph {et~al.}(2023)\citenamefont {Pan},
  \citenamefont {Kanitz}, \citenamefont {Debiossac}, \citenamefont {Le-Guen},
  \citenamefont {Rad},\ and\ \citenamefont {Roncin}}]{pan2023lateral}%
  \BibitemOpen
  \bibfield  {author} {\bibinfo {author} {\bibfnamefont {P.}~\bibnamefont
  {Pan}}, \bibinfo {author} {\bibfnamefont {C.}~\bibnamefont {Kanitz}},
  \bibinfo {author} {\bibfnamefont {M.}~\bibnamefont {Debiossac}}, \bibinfo
  {author} {\bibfnamefont {A.}~\bibnamefont {Le-Guen}}, \bibinfo {author}
  {\bibfnamefont {J.~N.}\ \bibnamefont {Rad}},\ and\ \bibinfo {author}
  {\bibfnamefont {P.}~\bibnamefont {Roncin}},\ }\bibfield  {title} {\bibinfo
  {title} {Lateral line profiles in fast-atom diffraction at surfaces},\
  }\href@noop {} {\bibfield  {journal} {\bibinfo  {journal} {Physical Review
  B}\ }\textbf {\bibinfo {volume} {108}},\ \bibinfo {pages} {035413} (\bibinfo
  {year} {2023})}\BibitemShut {NoStop}%
\bibitem [{sig()}]{sigma_phi}%
  \BibitemOpen
  \href@noop {} {}\bibinfo {note} {For instance, the width parameter for 4 keV
  Kr atoms was measured between 0.5$^\circ$ and 1.5$^\circ$ and modeled with a
  polynolial fit; $\sigma_\phi$(mdeg)=8+30 $\theta_i$+14 $\theta_i^2$ ($\theta$
  in deg). For 4 keV Xe, we convoluted the LG profile in eq.\ref{eq:LG} by the
  beam profile ($\sigma$=9 mdeg.) $\sigma_\phi$=-1.5+46 $\theta_i$+21
  $\theta_i^2$.}\BibitemShut {Stop}%
\bibitem [{Spe()}]{Specular}%
  \BibitemOpen
  \href@noop {} {}\bibinfo {note} {\textit{Stricto sensu}, there is only one
  specular spot corresponding to $\theta_f=-\theta_i$ and $\phi_f=\phi_i$, we
  call here specular circle the circle corresponding to
  $|\theta_f|=|\theta_i|$.}\BibitemShut {Stop}%
\bibitem [{\citenamefont {Momeni}\ \emph {et~al.}(2010)\citenamefont {Momeni},
  \citenamefont {Soulisse}, \citenamefont {Rousseau}, \citenamefont
  {Khemliche},\ and\ \citenamefont {Roncin}}]{momeni2010grazing}%
  \BibitemOpen
  \bibfield  {author} {\bibinfo {author} {\bibfnamefont {A.}~\bibnamefont
  {Momeni}}, \bibinfo {author} {\bibfnamefont {P.}~\bibnamefont {Soulisse}},
  \bibinfo {author} {\bibfnamefont {P.}~\bibnamefont {Rousseau}}, \bibinfo
  {author} {\bibfnamefont {H.}~\bibnamefont {Khemliche}},\ and\ \bibinfo
  {author} {\bibfnamefont {P.}~\bibnamefont {Roncin}},\ }\bibfield  {title}
  {\bibinfo {title} {Grazing incidence fast atom diffraction ({GIFAD}): Doing
  {RHEED} with atoms},\ }\href {https://doi.org/10.1380/ejssnt.2010.101}
  {\bibfield  {journal} {\bibinfo  {journal} {e-Journal of Surface Science and
  Nanotechnology}\ }\textbf {\bibinfo {volume} {8}},\ \bibinfo {pages} {101}
  (\bibinfo {year} {2010})}\BibitemShut {NoStop}%
\bibitem [{\citenamefont {Young}(2016)}]{young2016bakerian}%
  \BibitemOpen
  \bibfield  {author} {\bibinfo {author} {\bibfnamefont {T.}~\bibnamefont
  {Young}},\ }\href@noop {} {\bibinfo {title} {The bakerian lecture::
  Experiments and calculations relative to physical optics}} (\bibinfo {year}
  {2016})\BibitemShut {NoStop}%
\bibitem [{\citenamefont {Laven}(2004)}]{laven2004simulation}%
  \BibitemOpen
  \bibfield  {author} {\bibinfo {author} {\bibfnamefont {P.}~\bibnamefont
  {Laven}},\ }\bibfield  {title} {\bibinfo {title} {Simulation of rainbows,
  coronas and glories using mie theory and the debye series},\ }\href@noop {}
  {\bibfield  {journal} {\bibinfo  {journal} {Journal of Quantitative
  Spectroscopy and Radiative Transfer}\ }\textbf {\bibinfo {volume} {89}},\
  \bibinfo {pages} {257} (\bibinfo {year} {2004})}\BibitemShut {NoStop}%
\bibitem [{\citenamefont {Chow}\ and\ \citenamefont
  {Thompson}(1976)}]{chow1976bound}%
  \BibitemOpen
  \bibfield  {author} {\bibinfo {author} {\bibfnamefont {H.}~\bibnamefont
  {Chow}}\ and\ \bibinfo {author} {\bibfnamefont {E.}~\bibnamefont
  {Thompson}},\ }\bibfield  {title} {\bibinfo {title} {Bound state resonances
  in atom-solid scattering},\ }\href@noop {} {\bibfield  {journal} {\bibinfo
  {journal} {Surface Science}\ }\textbf {\bibinfo {volume} {59}},\ \bibinfo
  {pages} {225} (\bibinfo {year} {1976})}\BibitemShut {NoStop}%
\bibitem [{\citenamefont {Miret-Artes}\ and\ \citenamefont
  {Pollak}(2017)}]{miret2017scattering}%
  \BibitemOpen
  \bibfield  {author} {\bibinfo {author} {\bibfnamefont {S.}~\bibnamefont
  {Miret-Artes}}\ and\ \bibinfo {author} {\bibfnamefont {E.}~\bibnamefont
  {Pollak}},\ }\bibfield  {title} {\bibinfo {title} {Scattering of {He} atoms
  from a microstructured grating: quantum reflection probabilities and
  diffraction patterns},\ }\href@noop {} {\bibfield  {journal} {\bibinfo
  {journal} {The Journal of Physical Chemistry Letters}\ }\textbf {\bibinfo
  {volume} {8}},\ \bibinfo {pages} {1009} (\bibinfo {year} {2017})}\BibitemShut
  {NoStop}%
\bibitem [{\citenamefont {Garibaldi}\ \emph {et~al.}(1975)\citenamefont
  {Garibaldi}, \citenamefont {Levi}, \citenamefont {Spadacini},\ and\
  \citenamefont {Tommei}}]{garibaldi1975quantum}%
  \BibitemOpen
  \bibfield  {author} {\bibinfo {author} {\bibfnamefont {U.}~\bibnamefont
  {Garibaldi}}, \bibinfo {author} {\bibfnamefont {A.}~\bibnamefont {Levi}},
  \bibinfo {author} {\bibfnamefont {R.}~\bibnamefont {Spadacini}},\ and\
  \bibinfo {author} {\bibfnamefont {G.}~\bibnamefont {Tommei}},\ }\bibfield
  {title} {\bibinfo {title} {Quantum theory of atom-surface scattering:
  diffraction and rainbow},\ }\href@noop {} {\bibfield  {journal} {\bibinfo
  {journal} {Surface Science}\ }\textbf {\bibinfo {volume} {48}},\ \bibinfo
  {pages} {649} (\bibinfo {year} {1975})}\BibitemShut {NoStop}%
\bibitem [{\citenamefont {Sch\"uller}\ and\ \citenamefont
  {Winter}(2008)}]{Schueller2008Supernumerary}%
  \BibitemOpen
  \bibfield  {author} {\bibinfo {author} {\bibfnamefont {A.}~\bibnamefont
  {Sch\"uller}}\ and\ \bibinfo {author} {\bibfnamefont {H.}~\bibnamefont
  {Winter}},\ }\bibfield  {title} {\bibinfo {title} {Supernumerary rainbows in
  the angular distribution of scattered projectiles for grazing collisions of
  fast atoms with a {LiF}(001) surface},\ }\href
  {https://doi.org/10.1103/PhysRevLett.100.097602} {\bibfield  {journal}
  {\bibinfo  {journal} {Phys. Rev. Lett.}\ }\textbf {\bibinfo {volume} {100}},\
  \bibinfo {pages} {097602} (\bibinfo {year} {2008})}\BibitemShut {NoStop}%
\bibitem [{\citenamefont {Sch{\"u}ller}\ \emph {et~al.}(2005)\citenamefont
  {Sch{\"u}ller}, \citenamefont {Wethekam}, \citenamefont {Mertens},
  \citenamefont {Maass}, \citenamefont {Winter},\ and\ \citenamefont
  {G{\"a}rtner}}]{schuller2005interatomic}%
  \BibitemOpen
  \bibfield  {author} {\bibinfo {author} {\bibfnamefont {A.}~\bibnamefont
  {Sch{\"u}ller}}, \bibinfo {author} {\bibfnamefont {S.}~\bibnamefont
  {Wethekam}}, \bibinfo {author} {\bibfnamefont {A.}~\bibnamefont {Mertens}},
  \bibinfo {author} {\bibfnamefont {K.}~\bibnamefont {Maass}}, \bibinfo
  {author} {\bibfnamefont {H.}~\bibnamefont {Winter}},\ and\ \bibinfo {author}
  {\bibfnamefont {K.}~\bibnamefont {G{\"a}rtner}},\ }\bibfield  {title}
  {\bibinfo {title} {Interatomic potentials from rainbow scattering of kev
  noble gas atoms under axial surface channeling},\ }\href@noop {} {\bibfield
  {journal} {\bibinfo  {journal} {NIM-B}\ }\textbf {\bibinfo {volume} {230}},\
  \bibinfo {pages} {172} (\bibinfo {year} {2005})}\BibitemShut {NoStop}%
\bibitem [{\citenamefont {Specht}\ \emph {et~al.}(2011)\citenamefont {Specht},
  \citenamefont {Busch}, \citenamefont {Seifert}, \citenamefont {Sch{\"u}ller},
  \citenamefont {Winter}, \citenamefont {G{\"a}rtner}, \citenamefont
  {W{\l}odarczyk}, \citenamefont {Sierka},\ and\ \citenamefont
  {Sauer}}]{specht2011rainbow}%
  \BibitemOpen
  \bibfield  {author} {\bibinfo {author} {\bibfnamefont {U.}~\bibnamefont
  {Specht}}, \bibinfo {author} {\bibfnamefont {M.}~\bibnamefont {Busch}},
  \bibinfo {author} {\bibfnamefont {J.}~\bibnamefont {Seifert}}, \bibinfo
  {author} {\bibfnamefont {A.}~\bibnamefont {Sch{\"u}ller}}, \bibinfo {author}
  {\bibfnamefont {H.}~\bibnamefont {Winter}}, \bibinfo {author} {\bibfnamefont
  {K.}~\bibnamefont {G{\"a}rtner}}, \bibinfo {author} {\bibfnamefont
  {R.}~\bibnamefont {W{\l}odarczyk}}, \bibinfo {author} {\bibfnamefont
  {M.}~\bibnamefont {Sierka}},\ and\ \bibinfo {author} {\bibfnamefont
  {J.}~\bibnamefont {Sauer}},\ }\bibfield  {title} {\bibinfo {title} {Rainbow
  scattering under axial surface channeling from a {KCl} (001) surface},\
  }\href@noop {} {\bibfield  {journal} {\bibinfo  {journal} {Phys. Rev. B.}\
  }\textbf {\bibinfo {volume} {84}},\ \bibinfo {pages} {125440} (\bibinfo
  {year} {2011})}\BibitemShut {NoStop}%
\bibitem [{cor()}]{corrug_100}%
  \BibitemOpen
  \href@noop {} {}\bibinfo {note} {Note that for a cosine corrugation function
  where $I_m=J_m^2(k_\perp z_c)$, $\sigma_m$ is analytic:
  $\sigma_m=\sqrt{2}\,k_\perp\,z_c$ \cite{Pollak2024GIFAD}.}\BibitemShut
  {Stop}%
\bibitem [{\citenamefont {Pollak}\ and\ \citenamefont
  {Miret-Art{\'e}s}(2015)}]{pollak2015second}%
  \BibitemOpen
  \bibfield  {author} {\bibinfo {author} {\bibfnamefont {E.}~\bibnamefont
  {Pollak}}\ and\ \bibinfo {author} {\bibfnamefont {S.}~\bibnamefont
  {Miret-Art{\'e}s}},\ }\bibfield  {title} {\bibinfo {title} {Second-order
  semiclassical perturbation theory for diffractive scattering from a
  surface},\ }\href {https://doi.org/10.1021/jp509500v} {\bibfield  {journal}
  {\bibinfo  {journal} {J. Phys. Chem. C.}\ }\textbf {\bibinfo {volume}
  {119}},\ \bibinfo {pages} {14532} (\bibinfo {year} {2015})}\BibitemShut
  {NoStop}%
\bibitem [{\citenamefont {Allison}\ \emph {et~al.}(2022)\citenamefont
  {Allison}, \citenamefont {Miret-Art{\'e}s},\ and\ \citenamefont
  {Pollak}}]{allison2022perturbation}%
  \BibitemOpen
  \bibfield  {author} {\bibinfo {author} {\bibfnamefont {W.}~\bibnamefont
  {Allison}}, \bibinfo {author} {\bibfnamefont {S.}~\bibnamefont
  {Miret-Art{\'e}s}},\ and\ \bibinfo {author} {\bibfnamefont {E.}~\bibnamefont
  {Pollak}},\ }\bibfield  {title} {\bibinfo {title} {Perturbation theory of
  scattering for grazing-incidence fast-atom diffraction},\ }\href@noop {}
  {\bibfield  {journal} {\bibinfo  {journal} {Phys. Chem. Chem. Phys}\ }\textbf
  {\bibinfo {volume} {24}},\ \bibinfo {pages} {15851} (\bibinfo {year}
  {2022})}\BibitemShut {NoStop}%
\bibitem [{Off()}]{Off_axis}%
  \BibitemOpen
  \href@noop {} {}\bibinfo {note} {As soon as the beam direction is not aligned
  with the crystal axis, perturbation theory\cite{pollak2015second} clearly
  outlines the crucial role of the decay-range of the electron density also
  responsible for softwall corrections.}\BibitemShut {Stop}%
\bibitem [{\citenamefont {Miret-Art{\'e}s}\ and\ \citenamefont
  {Pollak}(2012)}]{miret2012classical}%
  \BibitemOpen
  \bibfield  {author} {\bibinfo {author} {\bibfnamefont {S.}~\bibnamefont
  {Miret-Art{\'e}s}}\ and\ \bibinfo {author} {\bibfnamefont {E.}~\bibnamefont
  {Pollak}},\ }\bibfield  {title} {\bibinfo {title} {Classical theory of
  atom--surface scattering: The rainbow effect},\ }\href
  {https://doi.org/10.1016/j.surfrep.2012.03.001} {\bibfield  {journal}
  {\bibinfo  {journal} {Surface Science Reports}\ }\textbf {\bibinfo {volume}
  {67}},\ \bibinfo {pages} {161} (\bibinfo {year} {2012})}\BibitemShut
  {NoStop}%
\bibitem [{\citenamefont {Meyer}(2015)}]{Meyer_thesis}%
  \BibitemOpen
  \bibfield  {author} {\bibinfo {author} {\bibfnamefont {E.}~\bibnamefont
  {Meyer}},\ }\href {https://doi.org/10.18452/17442} {\bibinfo {type}
  {Habilitattionsschrift {PhD} dissertation}},\ \bibinfo  {school} {Humboldt
  Universität Berlin, Germany} (\bibinfo {year} {2015})\BibitemShut {NoStop}%
\bibitem [{\citenamefont {Bocan}\ and\ \citenamefont
  {Gravielle}(2018)}]{bocan2018gifad}%
  \BibitemOpen
  \bibfield  {author} {\bibinfo {author} {\bibfnamefont {G.~A.}\ \bibnamefont
  {Bocan}}\ and\ \bibinfo {author} {\bibfnamefont {M.~S.}\ \bibnamefont
  {Gravielle}},\ }\bibfield  {title} {\bibinfo {title} {Gifad for {he/KCl}
  (001). structure in the pattern for [110] incidence as a measure of the
  projectile-cation interaction},\ }\href
  {https://doi.org/10.1016/j.nimb.2018.02.004} {\bibfield  {journal} {\bibinfo
  {journal} {NIM-B}\ }\textbf {\bibinfo {volume} {421}},\ \bibinfo {pages} {1}
  (\bibinfo {year} {2018})}\BibitemShut {NoStop}%
\bibitem [{\citenamefont {del Cueto}\ \emph {et~al.}(2020)\citenamefont {del
  Cueto}, \citenamefont {Muzas}, \citenamefont {Mart{\'\i}n},\ and\
  \citenamefont {D{\'\i}az}}]{del2020accurate}%
  \BibitemOpen
  \bibfield  {author} {\bibinfo {author} {\bibfnamefont {M.}~\bibnamefont {del
  Cueto}}, \bibinfo {author} {\bibfnamefont {A.}~\bibnamefont {Muzas}},
  \bibinfo {author} {\bibfnamefont {F.}~\bibnamefont {Mart{\'\i}n}},\ and\
  \bibinfo {author} {\bibfnamefont {C.}~\bibnamefont {D{\'\i}az}},\ }\bibfield
  {title} {\bibinfo {title} {Accurate simulations of atomic diffractive
  scattering from {KCl} (001) under fast grazing incidence conditions},\
  }\href@noop {} {\bibfield  {journal} {\bibinfo  {journal} {NIM-B}\ }\textbf
  {\bibinfo {volume} {476}},\ \bibinfo {pages} {1} (\bibinfo {year}
  {2020})}\BibitemShut {NoStop}%
\bibitem [{\citenamefont {Beeby}(1971)}]{Beeby1971}%
  \BibitemOpen
  \bibfield  {author} {\bibinfo {author} {\bibfnamefont {J.~L.}\ \bibnamefont
  {Beeby}},\ }\bibfield  {title} {\bibinfo {title} {The scattering of helium
  atoms from surfaces},\ }\href@noop {} {\bibfield  {journal} {\bibinfo
  {journal} {Journal of Physics C.}\ }\textbf {\bibinfo {volume} {4}},\
  \bibinfo {pages} {L359} (\bibinfo {year} {1971})}\BibitemShut {NoStop}%
\bibitem [{\citenamefont {Far\'{\i}as}\ and\ \citenamefont
  {Rieder}(1998)}]{Farias1998}%
  \BibitemOpen
  \bibfield  {author} {\bibinfo {author} {\bibfnamefont {D.}~\bibnamefont
  {Far\'{\i}as}}\ and\ \bibinfo {author} {\bibfnamefont {K.}~\bibnamefont
  {Rieder}},\ }\bibfield  {title} {\bibinfo {title} {Atomic beam diffraction
  from solid surfaces},\ }\href {https://doi.org/10.1007/s003390100847}
  {\bibfield  {journal} {\bibinfo  {journal} {Rep. Prog. Phys.}\ }\textbf
  {\bibinfo {volume} {61}},\ \bibinfo {pages} {1575} (\bibinfo {year}
  {1998})}\BibitemShut {NoStop}%
\bibitem [{\citenamefont {Hoinkes}(1980)}]{hoinkes1980physical}%
  \BibitemOpen
  \bibfield  {author} {\bibinfo {author} {\bibfnamefont {H.}~\bibnamefont
  {Hoinkes}},\ }\bibfield  {title} {\bibinfo {title} {The physical interaction
  potential of gas atoms with single-crystal surfaces, determined from
  gas-surface diffraction experiments},\ }\href@noop {} {\bibfield  {journal}
  {\bibinfo  {journal} {Reviews of Modern Physics}\ }\textbf {\bibinfo {volume}
  {52}},\ \bibinfo {pages} {933} (\bibinfo {year} {1980})}\BibitemShut
  {NoStop}%
\bibitem [{sam()}]{same_D}%
  \BibitemOpen
  \href@noop {} {}\bibinfo {note} {When comparing calculations or models to
  diffraction, the attractive part $D$ is sometime used as a adjusting
  parameter but then, $D$ should be the same along all directions. The
  conclusion in the VdW force are not important\cite{bocan2021dynamical} along
  [100] but important \cite{bocan2020anomalous} along the [110] direction to
  explain an anomalous behavior should be argumented. Note alose that these
  work contain no reference to refraction effect
  \cite{Farias1998,Mantovani,debiossac2020refraction} or the Beeby correction
  \cite{Beeby1971}.}\BibitemShut {Stop}%
\bibitem [{\citenamefont {Ziegler}\ and\ \citenamefont {Biersack}()}]{zbl}%
  \BibitemOpen
  \bibfield  {author} {\bibinfo {author} {\bibfnamefont {J.~F.}\ \bibnamefont
  {Ziegler}}\ and\ \bibinfo {author} {\bibfnamefont {J.~P.}\ \bibnamefont
  {Biersack}},\ }\bibinfo {title} {The stopping and range of ions in
  matter}\BibitemShut {NoStop}%
\bibitem [{\citenamefont {Lecoffre}\ \emph {et~al.}(2025)\citenamefont
  {Lecoffre}, \citenamefont {Hadi}, \citenamefont {Bruneau}, \citenamefont
  {Garcion}, \citenamefont {Fabre}, \citenamefont {Charron}, \citenamefont
  {Gaaloul}, \citenamefont {Dutier},\ and\ \citenamefont
  {Bouton}}]{lecoffre2025measurement}%
  \BibitemOpen
  \bibfield  {author} {\bibinfo {author} {\bibfnamefont {J.}~\bibnamefont
  {Lecoffre}}, \bibinfo {author} {\bibfnamefont {A.}~\bibnamefont {Hadi}},
  \bibinfo {author} {\bibfnamefont {M.}~\bibnamefont {Bruneau}}, \bibinfo
  {author} {\bibfnamefont {C.}~\bibnamefont {Garcion}}, \bibinfo {author}
  {\bibfnamefont {N.}~\bibnamefont {Fabre}}, \bibinfo {author} {\bibfnamefont
  {E.}~\bibnamefont {Charron}}, \bibinfo {author} {\bibfnamefont
  {N.}~\bibnamefont {Gaaloul}}, \bibinfo {author} {\bibfnamefont
  {G.}~\bibnamefont {Dutier}},\ and\ \bibinfo {author} {\bibfnamefont
  {Q.}~\bibnamefont {Bouton}},\ }\bibfield  {title} {\bibinfo {title}
  {Measurement of {Casimir-Polder} interaction for slow atoms through a
  material grating},\ }\href@noop {} {\bibfield  {journal} {\bibinfo  {journal}
  {Physical Review Research}\ }\textbf {\bibinfo {volume} {7}},\ \bibinfo
  {pages} {013232} (\bibinfo {year} {2025})}\BibitemShut {NoStop}%
\bibitem [{\citenamefont {Debiossac}\ \emph
  {et~al.}(2014{\natexlab{b}})\citenamefont {Debiossac}, \citenamefont
  {Zugarramurdi}, \citenamefont {Khemliche}, \citenamefont {Roncin},
  \citenamefont {Borisov}, \citenamefont {Momeni}, \citenamefont {Atkinson},
  \citenamefont {Eddrief}, \citenamefont {Finocchi},\ and\ \citenamefont
  {Etgens}}]{Debiossac_PRB_2014}%
  \BibitemOpen
  \bibfield  {author} {\bibinfo {author} {\bibfnamefont {M.}~\bibnamefont
  {Debiossac}}, \bibinfo {author} {\bibfnamefont {A.}~\bibnamefont
  {Zugarramurdi}}, \bibinfo {author} {\bibfnamefont {H.}~\bibnamefont
  {Khemliche}}, \bibinfo {author} {\bibfnamefont {P.}~\bibnamefont {Roncin}},
  \bibinfo {author} {\bibfnamefont {A.~G.}\ \bibnamefont {Borisov}}, \bibinfo
  {author} {\bibfnamefont {A.}~\bibnamefont {Momeni}}, \bibinfo {author}
  {\bibfnamefont {P.}~\bibnamefont {Atkinson}}, \bibinfo {author}
  {\bibfnamefont {M.}~\bibnamefont {Eddrief}}, \bibinfo {author} {\bibfnamefont
  {F.}~\bibnamefont {Finocchi}},\ and\ \bibinfo {author} {\bibfnamefont
  {V.~H.}\ \bibnamefont {Etgens}},\ }\bibfield  {title} {\bibinfo {title}
  {Combined experimental and theoretical study of fast atom diffraction on the
  ${\ensuremath{\beta}}_{2}(2\ifmmode\times\else\texttimes\fi{}4)$
  reconstructed {GaAs}(001) surface},\ }\href
  {https://doi.org/10.1103/PhysRevB.90.155308} {\bibfield  {journal} {\bibinfo
  {journal} {Phys. Rev. B}\ }\textbf {\bibinfo {volume} {90}},\ \bibinfo
  {pages} {155308} (\bibinfo {year} {2014}{\natexlab{b}})}\BibitemShut
  {NoStop}%
\bibitem [{\citenamefont {Debiossac}\ \emph {et~al.}(2016)\citenamefont
  {Debiossac}, \citenamefont {Zugarramurdi}, \citenamefont {Mu}, \citenamefont
  {Lunca-Popa}, \citenamefont {Mayne},\ and\ \citenamefont
  {Roncin}}]{Debiossac_PRB_2016}%
  \BibitemOpen
  \bibfield  {author} {\bibinfo {author} {\bibfnamefont {M.}~\bibnamefont
  {Debiossac}}, \bibinfo {author} {\bibfnamefont {A.}~\bibnamefont
  {Zugarramurdi}}, \bibinfo {author} {\bibfnamefont {Z.}~\bibnamefont {Mu}},
  \bibinfo {author} {\bibfnamefont {P.}~\bibnamefont {Lunca-Popa}}, \bibinfo
  {author} {\bibfnamefont {A.~J.}\ \bibnamefont {Mayne}},\ and\ \bibinfo
  {author} {\bibfnamefont {P.}~\bibnamefont {Roncin}},\ }\bibfield  {title}
  {\bibinfo {title} {Helium diffraction on {SiC} grown graphene: Qualitative
  and quantitative descriptions with the hard-corrugated-wall model},\ }\href
  {https://doi.org/10.1103/PhysRevB.94.205403} {\bibfield  {journal} {\bibinfo
  {journal} {Phys. Rev. B}\ }\textbf {\bibinfo {volume} {94}},\ \bibinfo
  {pages} {205403} (\bibinfo {year} {2016})}\BibitemShut {NoStop}%
\bibitem [{\citenamefont {Bocan}\ \emph {et~al.}(2021)\citenamefont {Bocan},
  \citenamefont {Breiss}, \citenamefont {Szilasi}, \citenamefont {Momeni},
  \citenamefont {Staicu~Casagrande}, \citenamefont {Sanchez}, \citenamefont
  {Gravielle},\ and\ \citenamefont {Khemliche}}]{bocan2021dynamical}%
  \BibitemOpen
  \bibfield  {author} {\bibinfo {author} {\bibfnamefont {G.~A.}\ \bibnamefont
  {Bocan}}, \bibinfo {author} {\bibfnamefont {H.}~\bibnamefont {Breiss}},
  \bibinfo {author} {\bibfnamefont {S.}~\bibnamefont {Szilasi}}, \bibinfo
  {author} {\bibfnamefont {A.}~\bibnamefont {Momeni}}, \bibinfo {author}
  {\bibfnamefont {E.}~\bibnamefont {Staicu~Casagrande}}, \bibinfo {author}
  {\bibfnamefont {E.~A.}\ \bibnamefont {Sanchez}}, \bibinfo {author}
  {\bibfnamefont {M.~S.}\ \bibnamefont {Gravielle}},\ and\ \bibinfo {author}
  {\bibfnamefont {H.}~\bibnamefont {Khemliche}},\ }\bibfield  {title} {\bibinfo
  {title} {Dynamical effects as a window into van der waals interactions in
  grazing-incidence fast he-atom diffraction from kcl (001)},\ }\href@noop {}
  {\bibfield  {journal} {\bibinfo  {journal} {Physical Review B}\ }\textbf
  {\bibinfo {volume} {104}},\ \bibinfo {pages} {235401} (\bibinfo {year}
  {2021})}\BibitemShut {NoStop}%
\bibitem [{\citenamefont {Bocan}\ \emph {et~al.}(2020)\citenamefont {Bocan},
  \citenamefont {Breiss}, \citenamefont {Szilasi}, \citenamefont {Momeni},
  \citenamefont {Casagrande}, \citenamefont {Gravielle}, \citenamefont
  {Sanchez},\ and\ \citenamefont {Khemliche}}]{bocan2020anomalous}%
  \BibitemOpen
  \bibfield  {author} {\bibinfo {author} {\bibfnamefont {G.~A.}\ \bibnamefont
  {Bocan}}, \bibinfo {author} {\bibfnamefont {H.}~\bibnamefont {Breiss}},
  \bibinfo {author} {\bibfnamefont {S.}~\bibnamefont {Szilasi}}, \bibinfo
  {author} {\bibfnamefont {A.}~\bibnamefont {Momeni}}, \bibinfo {author}
  {\bibfnamefont {E.~S.}\ \bibnamefont {Casagrande}}, \bibinfo {author}
  {\bibfnamefont {M.~S.}\ \bibnamefont {Gravielle}}, \bibinfo {author}
  {\bibfnamefont {E.~A.}\ \bibnamefont {Sanchez}},\ and\ \bibinfo {author}
  {\bibfnamefont {H.}~\bibnamefont {Khemliche}},\ }\bibfield  {title} {\bibinfo
  {title} {Anomalous kcl (001) surface corrugation from fast he diffraction at
  very grazing incidence},\ }\href@noop {} {\bibfield  {journal} {\bibinfo
  {journal} {Physical Review Letters}\ }\textbf {\bibinfo {volume} {125}},\
  \bibinfo {pages} {096101} (\bibinfo {year} {2020})}\BibitemShut {NoStop}%
\bibitem [{\citenamefont {Mantovani}\ and\ \citenamefont
  {Manson}(1982)}]{Mantovani}%
  \BibitemOpen
  \bibfield  {author} {\bibinfo {author} {\bibfnamefont {J.~G.}\ \bibnamefont
  {Mantovani}}\ and\ \bibinfo {author} {\bibfnamefont {J.~R.}\ \bibnamefont
  {Manson}},\ }\bibfield  {title} {\bibinfo {title} {Simple approach to
  refraction effects in atom-surface scattering},\ }\href
  {https://doi.org/10.1016/0039-6028(82)90145-5} {\bibfield  {journal}
  {\bibinfo  {journal} {Surface Science}\ }\textbf {\bibinfo {volume} {120}},\
  \bibinfo {pages} {L487} (\bibinfo {year} {1982})}\BibitemShut {NoStop}%
\end{thebibliography}%

\end{document}